\documentclass[11pt, a4paper]{article}

\usepackage{jheppub}

\DeclareFontFamily{U}{MnSymbolC}{}
\DeclareSymbolFont{MnSyC}{U}{MnSymbolC}{m}{n}
\DeclareFontShape{U}{MnSymbolC}{m}{n}{
    <-6>  MnSymbolC5
   <6-7>  MnSymbolC6
   <7-8>  MnSymbolC7
   <8-9>  MnSymbolC8
   <9-10> MnSymbolC9
  <10-12> MnSymbolC10
  <12->   MnSymbolC12}{}
\DeclareMathSymbol{\intprod}{\mathbin}{MnSyC}{'270}

\newcommand{\del}{\partial}
\newcommand{\delb}{{\bar\partial}}

\newcommand{\vev}[1]{\langle #1 \rangle}

\newcommand{\Map}{\mathop{\mathrm{Map}}\nolimits}

\renewcommand{\Im}{\mathop{\mathrm{Im}}\nolimits}
\renewcommand{\Re}{\mathop{\mathrm{Re}}\nolimits}

\newcommand{\Tr}{\mathop{\mathrm{Tr}}\nolimits}

\newcommand{\OO}{\mathrm{O}}

\newcommand{\U}{\mathrm{U}}

\newcommand{\Z}{\mathbb{Z}}

\newcommand{\R}{\mathbb{R}}
\newcommand{\C}{\mathbb{C}}

\renewcommand{\P}{\mathbb{CP}}

\let\nc\newcommand
\let\renc\renewcommand
\nc{\wbar}{\overline}
\let\td\tilde
\let\wtd\widetilde
\let\wht\widehat
\let\mcl\mathcal

\nc{\ab}{{\bar{a}}} \nc{\at}{\tilde{a}} \nc{\ah}{\hat{a}}
\nc{\bb}{{\bar{b}}} \nc{\bt}{\tilde{b}} \nc{\bh}{\hat{b}}
\nc{\cb}{{\bar{c}}} \nc{\ct}{\tilde{c}} 
\nc{\db}{{\bar{d}}} \nc{\dt}{\tilde{d}} \renc{\dh}{\hat{d}}
\nc{\eb}{{\bar{e}}} \nc{\et}{\tilde{e}} \nc{\eh}{\hat{e}}
\nc{\fb}{{\bar{f}}} \nc{\ft}{\tilde{f}} \nc{\fh}{\hat{f}}
\nc{\gb}{{\bar{g}}} \nc{\gt}{\tilde{g}} \nc{\gh}{\hat{g}}
\nc{\hb}{{\bar{h}}} \nc{\hh}{\hat{h}} 
\nc{\ib}{{\bar{\imath}}} \nc{\ih}{\hat{\imath}} 
\nc{\jb}{{\bar{\jmath}}} \nc{\jt}{\tilde{\jmath}} \nc{\jh}{\hat{\jmath}}
\nc{\kb}{{\bar{k}}} \nc{\kt}{\tilde{k}} \nc{\kh}{\hat{k}}
\nc{\lb}{{\bar{l}}} \nc{\lt}{\tilde{l}} \nc{\lh}{\hat{l}}
\nc{\mb}{{\bar{m}}} \nc{\mt}{\tilde{m}} \nc{\mh}{\hat{m}}
\nc{\nb}{{\bar{n}}} \nc{\nt}{\tilde{n}} \nc{\nh}{\hat{n}}
\nc{\ob}{{\bar{o}}} \nc{\ot}{\tilde{o}} \nc{\oh}{\hat{o}}
\nc{\pb}{{\bar{p}}} \nc{\pt}{\tilde{p}} \nc{\ph}{\hat{p}}
\nc{\qb}{{\bar{q}}} \nc{\qt}{\tilde{q}} \nc{\qh}{\hat{q}}
\nc{\rb}{{\bar{r}}} \nc{\rt}{\tilde{r}} \nc{\rh}{\hat{r}}
\renc{\sb}{{\bar{s}}} \nc{\st}{\tilde{s}} \nc{\sh}{\hat{s}}
\nc{\tb}{{\bar{t}}} \renc{\th}{\hat{t}} 
\nc{\ub}{{\bar{u}}} \nc{\ut}{\tilde{u}} \nc{\uh}{\hat{u}}
\nc{\vb}{{\bar{v}}} \nc{\vt}{\tilde{v}} \nc{\vh}{\hat{v}}
\nc{\wb}{{\bar{w}}} \nc{\wt}{\tilde{w}} \nc{\wh}{\hat{w}}
\nc{\xb}{{\bar{x}}} \nc{\xt}{\tilde{x}} \nc{\xh}{\hat{x}}
\nc{\yb}{{\bar{y}}} \nc{\yt}{\tilde{y}} \nc{\yh}{\hat{y}}
\nc{\zb}{{\bar{z}}} \nc{\zt}{\tilde{z}} \nc{\zh}{\hat{z}}

\nc{\Ab}{{\wbar{A}}} \nc{\At}{{\wtd{A}}} \nc{\Ah}{{\wht{A}}}
\nc{\Bb}{{\wbar{B}}} \nc{\Bt}{{\wtd{B}}} \nc{\Bh}{{\wht{B}}}
\nc{\Cb}{{\wbar{C}}} \nc{\Ct}{{\wtd{C}}} \nc{\Ch}{{\wht{C}}}
\nc{\Db}{{\wbar{D}}} \nc{\Dt}{{\wtd{D}}} \nc{\Dh}{{\wht{D}}}
\nc{\Eb}{{\wbar{E}}} \nc{\Et}{{\wtd{E}}} \nc{\Eh}{{\wht{E}}}
\nc{\Fb}{{\wbar{F}}} \nc{\Ft}{{\wtd{F}}} \nc{\Fh}{{\wht{F}}}
\nc{\Gb}{{\wbar{G}}} \nc{\Gt}{{\wtd{G}}} \nc{\Gh}{{\wht{G}}}
\nc{\Hb}{{\wbar{H}}} \nc{\Ht}{{\wtd{H}}} \nc{\Hh}{{\wht{H}}}
\nc{\Ib}{{\bar{I}}} \nc{\It}{{\wtd{I}}} \nc{\Ih}{{\wht{I}}}
\nc{\Jb}{{\bar{J}}} \nc{\Jt}{{\wtd{J}}} \nc{\Jh}{{\wht{J}}}
\nc{\Kb}{{\wbar{K}}} \nc{\Kt}{{\wtd{K}}} \nc{\Kh}{{\wht{K}}}
\nc{\Lb}{{\wbar{L}}} \nc{\Lt}{{\wtd{L}}} \nc{\Lh}{{\wht{L}}}
\nc{\Mb}{{\wbar{M}}} \nc{\Mt}{{\wtd{M}}} \nc{\Mh}{{\wht{M}}}
\nc{\Nb}{{\wbar{N}}} \nc{\Nt}{{\wtd{N}}} \nc{\Nh}{{\wht{N}}}
\nc{\Ob}{{\wbar{O}}} \nc{\Ot}{{\wtd{O}}} \nc{\Oh}{{\wht{O}}}
\nc{\Pb}{{\wbar{P}}} \nc{\Pt}{{\wtd{P}}} \nc{\Ph}{{\wht{P}}}
\nc{\Qb}{{\wbar{Q}}} \nc{\Qt}{{\wtd{Q}}} \nc{\Qh}{{\wht{Q}}}
\nc{\Rb}{{\wbar{R}}} \nc{\Rt}{{\wtd{R}}} \nc{\Rh}{{\wht{R}}}
\nc{\Sb}{{\wbar{S}}} \nc{\St}{{\wtd{S}}} \nc{\Sh}{{\wht{S}}}
\nc{\Tb}{{\wbar{T}}} \nc{\Tt}{{\wtd{T}}} \nc{\Th}{{\wht{T}}}
\nc{\Ub}{{\wbar{U}}} \nc{\Ut}{{\wtd{U}}} \nc{\Uh}{{\wht{U}}}
\nc{\Vb}{{\wbar{V}}} \nc{\Vt}{{\wtd{V}}} \nc{\Vh}{{\wht{V}}}
\nc{\Wb}{{\wbar{W}}} \nc{\Wt}{{\wtd{W}}} \nc{\Wh}{{\wht{W}}}
\nc{\Xb}{{\wbar{X}}} \nc{\Xt}{{\wtd{X}}} \nc{\Xh}{{\wht{X}}}
\nc{\Yb}{{\wbar{Y}}} \nc{\Yt}{{\wtd{Y}}} \nc{\Yh}{{\wht{Y}}}
\nc{\Zb}{{\wbar{Z}}} \nc{\Zt}{{\wtd{Z}}} \nc{\Zh}{{\wht{Z}}}

\nc{\CA}{{\mcl{A}}} \nc{\CAb}{{\wbar{\CA}}} \nc{\CAt}{{\wtd{\CA}}} \nc{\CAh}{{\wht{\CA}}}
\nc{\CB}{{\mcl{B}}} \nc{\CBb}{{\wbar{\CB}}} \nc{\CBt}{{\wtd{\CB}}} \nc{\CBh}{{\wht{\CB}}}
\nc{\CC}{{\mcl{C}}} \nc{\CCb}{{\wbar{\CC}}} \nc{\CCt}{{\wtd{\CC}}} \nc{\CCh}{{\wht{\CC}}}
\nc{\cD}{{\mcl{D}}} \nc{\cDb}{{\wbar{\cD}}} \nc{\cDt}{{\wtd{\cC}}} \nc{\cDh}{{\wht{\cD}}}
\nc{\CE}{{\mcl{E}}} \nc{\CEb}{{\wbar{\CE}}} \nc{\CEt}{{\wtd{\CE}}} \nc{\CEh}{{\wht{\CE}}}
\nc{\CF}{{\mcl{F}}} \nc{\CFb}{{\wbar{\CF}}} \nc{\CFt}{{\wtd{\CF}}} \nc{\CFh}{{\wht{\CF}}}
\nc{\CG}{{\mcl{G}}} \nc{\CGb}{{\wbar{\CG}}} \nc{\CGt}{{\wtd{\CG}}} \nc{\CGh}{{\wht{\CG}}}
\nc{\CH}{{\mcl{H}}} \nc{\CHb}{{\wbar{\CH}}} \nc{\CHt}{{\wtd{\CH}}} \nc{\CHh}{{\wht{\CH}}}
\nc{\CI}{{\mcl{I}}} \nc{\CIb}{{\wbar{\CI}}} \nc{\CIt}{{\wtd{\CI}}} \nc{\CIh}{{\wht{\CI}}}
\nc{\CJ}{{\mcl{J}}} \nc{\CJb}{{\wbar{\CJ}}} \nc{\CJt}{{\wtd{\CJ}}} \nc{\CJh}{{\wht{\CJ}}}
\nc{\CK}{{\mcl{K}}} \nc{\CKb}{{\wbar{\CK}}} \nc{\CKt}{{\wtd{\CK}}} \nc{\CKh}{{\wht{\CK}}}
\nc{\CL}{{\mcl{L}}} \nc{\CLb}{{\wbar{\CL}}} \nc{\CLt}{{\wtd{\CL}}} \nc{\CLh}{{\wht{\CL}}}
\nc{\CM}{{\mcl{M}}} \nc{\CMb}{{\wbar{\CM}}} \nc{\CMt}{{\wtd{\CM}}} \nc{\CMh}{{\wht{\CM}}}
\nc{\CN}{{\mcl{N}}} \nc{\CNb}{{\wbar{\CN}}} \nc{\CNt}{{\wtd{\CN}}} \nc{\CNh}{{\wht{\CN}}}
\nc{\CO}{{\mcl{O}}} \nc{\COb}{{\wbar{\CO}}} \nc{\COt}{{\wtd{\CO}}} \nc{\COh}{{\wht{\CO}}}
\nc{\CP}{{\mcl{P}}} \nc{\CPb}{{\wbar{\CP}}} \nc{\CPt}{{\wtd{\CP}}} \nc{\CPh}{{\wht{\CP}}}
\nc{\CQ}{{\mcl{Q}}} \nc{\CQb}{{\wbar{\CQ}}} \nc{\CQt}{{\wtd{\CQ}}} \nc{\CQh}{{\wht{\CQ}}}
\nc{\CR}{{\mcl{R}}} \nc{\CRb}{{\wbar{\CR}}} \nc{\CRt}{{\wtd{\CR}}} \nc{\CRh}{{\wht{\CR}}}
\nc{\CS}{{\mcl{S}}} \nc{\CSb}{{\wbar{\CS}}} \nc{\CSt}{{\wtd{\CS}}} \nc{\CSh}{{\wht{\CS}}}
\nc{\CT}{{\mcl{T}}} \nc{\CTb}{{\wbar{\CT}}} \nc{\CTt}{{\wtd{\CT}}} \nc{\CTh}{{\wht{\CT}}}
\nc{\CU}{{\mcl{U}}} \nc{\CUb}{{\wbar{\CU}}} \nc{\CUt}{{\wtd{\CU}}} \nc{\CUh}{{\wht{\CU}}}
\nc{\CV}{{\mcl{V}}} \nc{\CVb}{{\wbar{\CV}}} \nc{\CVt}{{\wtd{\CV}}} \nc{\CVh}{{\wht{\CV}}}
\nc{\CW}{{\mcl{W}}} \nc{\CWb}{{\wbar{\CW}}} \nc{\CWt}{{\wtd{\CW}}} \nc{\CWh}{{\wht{\CW}}}
\nc{\CX}{{\mcl{X}}} \nc{\CXb}{{\wbar{\CX}}} \nc{\CXt}{{\wtd{\CX}}} \nc{\CXh}{{\wht{\CX}}}
\nc{\CY}{{\mcl{Y}}} \nc{\CYb}{{\wbar{\CY}}} \nc{\CYt}{{\wtd{\CY}}} \nc{\CYh}{{\wht{\CY}}}
\nc{\CZ}{{\mcl{Z}}} \nc{\CZb}{{\wbar{\CZ}}} \nc{\CZt}{{\wtd{\CZ}}} \nc{\CZh}{{\wht{\CZ}}}

\let\eps\epsilon
\let\ups\upsilon
\let\veps\varepsilon
\let\vtht\vartheta
\let\vtheta\vtht
\let\vsgm\varsigma
\let\vphi\varphi
\let\vrho\varrho

\nc{\alphab}{{\bar{\alpha}}} \nc{\alphat}{{\td{\alpha}}} \nc{\alphah}{{\hat{\alpha}}}
\nc{\betab}{{\bar{\beta}}}   \nc{\betat}{{\td{\beta}}}   \nc{\betah}{{\hat{\beta}}} 
\nc{\gammab}{{\bar{\gamma}}} \nc{\gammat}{{\td{\gamma}}} \nc{\gammah}{{\hat{\gamma}}} 
\nc{\deltab}{{\bar{\delta}}} \nc{\deltat}{{\td{\delta}}} \nc{\deltah}{{\hat{\delta}}} 
\nc{\epsilonb}{{\bar{\eps}}} \nc{\epsilont}{{\td{\eps}}} \nc{\epsilonh}{{\hat{\eps}}} 
\nc{\vepsb}{{\bar{\veps}}}   \nc{\vepst}{{\td{\veps}}}   \nc{\vepsh}{{\hat{\veps}}} 
\nc{\zetab}{{\bar{\zeta}}}   \nc{\zetat}{{\td{\zeta}}}   \nc{\zetah}{{\hat{\zeta}}} 
\nc{\etab}{{\bar{\eta}}}     \nc{\etat}{{\td{\eta}}}     \nc{\etah}{{\hat{\eta}}} 
\nc{\thetab}{{\bar{\theta}}} \nc{\thetat}{{\td{\theta}}} \nc{\thetah}{{\hat{\theta}}} 
\nc{\vthetab}{{\bar{\vtht}}} \nc{\vthetat}{{\td{\vtht}}} \nc{\vthetah}{{\hat{\vtht}}} 
\nc{\lambdab}{{\bar{\lambda}}} \nc{\lambdat}{{\td{\lambda}}} \nc{\lambdah}{{\hat{\lambda}}} 
\nc{\iotab}{{\bar{\iota}}}   \nc{\iotat}{{\td{\iota}}}   \nc{\iotah}{{\hat{\iota}}} 
\nc{\kappab}{{\bar{\kappa}}} \nc{\kappat}{{\td{\kappa}}} \nc{\kappah}{{\hat{\kappa}}} 
\nc{\lmdb}{{\bar{\lmd}}}     \nc{\lmdt}{{\td{\lmd}}}     \nc{\lmdh}{{\hat{\lmd}}} 
\nc{\mub}{{\bar{\mu}}}       \nc{\mut}{{\td{\mu}}}       \nc{\muh}{{\hat{\mu}}} 
\nc{\nub}{{\bar{\nu}}}       \nc{\nut}{{\td{\nu}}}       \nc{\nuh}{{\hat{\nu}}} 
\nc{\xib}{{\bar{\xi}}}       \nc{\xit}{{\td{\xi}}}       \nc{\xih}{{\hat{\xi}}} 
\nc{\pib}{{\bar{\pi}}}       \nc{\pit}{{\td{\pi}}}       \nc{\pih}{{\hat{\pi}}} 
\nc{\vpib}{{\bar{\vpi}}}     \nc{\vpit}{{\td{\vpi}}}     \nc{\vpih}{{\hat{\vpi}}} 
\nc{\rhob}{{\bar{\rho}}}     \nc{\rhot}{{\td{\rho}}}     \nc{\rhoh}{{\hat{\rho}}} 
\nc{\vrhob}{{\bar{\vrho}}}   \nc{\vrhot}{{\td{\vrho}}}   \nc{\vrhoh}{{\hat{\vrho}}} 
\nc{\sigmab}{{\bar{\sigma}}} \nc{\sigmat}{{\td{\sigma}}} \nc{\sigmah}{{\hat{\sigma}}} 
\nc{\vsigmab}{{\bar{\vsgm}}} \nc{\vsigmat}{{\td{\vsgm}}} \nc{\vsigmah}{{\hat{\vsgm}}} 
\nc{\taub}{{\bar{\tau}}}     \nc{\taut}{{\td{\tau}}}     \nc{\tauh}{{\hat{\tau}}} 
\nc{\upsb}{{\bar{\ups}}} \nc{\upst}{{\td{\ups}}} \nc{\upsh}{{\hat{\ups}}} 
\nc{\phib}{{\bar{\phi}}}     \nc{\phit}{{\td{\phi}}}     \nc{\phih}{{\hat{\phi}}} 
\nc{\varphib}{{\bar{\vphi}}}   \nc{\varphit}{{\td{\vphi}}}   \nc{\varphih}{{\hat{\vphi}}} 
\nc{\chib}{{\bar{\chi}}}     \nc{\chit}{{\td{\chi}}}     \nc{\chih}{{\hat{\chi}}} 
\nc{\psib}{{\bar{\psi}}}     \nc{\psit}{{\td{\psi}}}     \nc{\psih}{{\hat{\psi}}} 
\nc{\omegab}{{\bar{\omega}}} \nc{\omegat}{{\td{\omega}}} \nc{\omegah}{{\hat{\omega}}} 

\nc{\Gammab}{{\wbar{\Gamma}}}     \nc{\Gammat}{{\wtd{\Gamma}}}     \nc{\Gammah}{{\wht{\Gamma}}}
\nc{\Deltab}{{\wbar{\Delta}}}     \nc{\Deltat}{{\wtd{\Delta}}}     \nc{\Deltah}{{\wht{\Delta}}}
\nc{\Thetab}{{\wbar{\Theta}}}     \nc{\Thetat}{{\wtd{\Theta}}}     \nc{\Thetah}{{\wht{\Theta}}}
\nc{\Lambdab}{{\wbar{\Lambda}}}   \nc{\Lambdat}{{\wtd{\Lambda}}}   \nc{\Lambdah}{{\wht{\Lambda}}}
\nc{\Xib}{{\wbar{\Xi}}}           \nc{\Xit}{{\wtd{\Xi}}}           \nc{\Xih}{{\wht{\Xi}}}
\nc{\Pib}{{\wbar{\Pi}}}           \nc{\Pit}{{\wtd{\Pi}}}           \nc{\Pih}{{\wht{\Pi}}}
\nc{\Sigmab}{{\wbar{\Sigma}}}     \nc{\Sigmat}{{\wtd{\Sigma}}}     \nc{\Sigmah}{{\wht{\Sigma}}}
\nc{\Upsilonb}{{\wbar{\Upsilon}}} \nc{\Upsilont}{{\wtd{\Upsilon}}} \nc{\Upsilonh}{{\wht{\Upsilon}}}
\nc{\Phib}{{\wbar{\Phi}}}         \nc{\Phit}{{\wtd{\Phi}}}         \nc{\Phih}{{\wht{\Phi}}}
\nc{\Psib}{{\wbar{\Psi}}}         \nc{\Psit}{{\wtd{\Psi}}}         \nc{\Psih}{{\wht{\Psi}}}
\nc{\Omegab}{{\wbar{\Omega}}}     \nc{\Omegat}{{\wtd{\Omega}}}     \nc{\Omegah}{{\wht{\Omega}}}

\newcommand{\rmd}{\mathrm{d}}

\title{\texorpdfstring{$\boldsymbol\Omega$}{Omega}-deformation and quantization}

\author{Junya Yagi}

\emailAdd{junya.yagi@sissa.it}
\affiliation{International School for Advanced Studies (SISSA) \\
via Bonomea, 265, 34136 Trieste, Italy
\\
INFN, Sezione di Trieste,
via Valerio, 2, 34149 Trieste, Italy}

\abstract{We formulate a deformation of Rozansky-Witten theory
  analogous to the $\Omega$-deformation.  It is applicable when the
  target space $X$ is hyperk\"ahler and the spacetime is of the form
  $\R \times \Sigma$, with $\Sigma$ being a Riemann surface.  In the
  case that $\Sigma$ is a disk, the $\Omega$-deformed Rozansky-Witten
  theory quantizes a symplectic submanifold of $X$, thereby providing
  a new perspective on quantization.  As applications, we elucidate
  two phenomena in four-dimensional gauge theory from this point of
  view.  One is a correspondence between the $\Omega$-deformation and
  quantization of integrable systems.  The other concerns
  supersymmetric loop operators and quantization of the algebra of
  holomorphic functions on a hyperk\"ahler manifold.}


\arxivnumber{}

\newcommand{\TYb}{\overline{TY}}
\newcommand{\nablat}{\widetilde\nabla}

\begin{document}
\maketitle
\flushbottom

\section{Introduction}
\label{Intro}

The goal of this paper is to develop a new perspective on
quantization, from which some intriguing phenomena in four-dimensional
gauge theory may be naturally understood.

Specifically, the phenomena that we wish to understand are the
following.  At low energies, an $\CN = 2$ supersymmetric gauge theory
compactified on a circle $S^1$ is described by a three-dimensional
$\CN = 4$ supersymmetric sigma model \cite{Seiberg:1996nz}.  The
target space $\CM$ of the sigma model is a hyperk\"ahler manifold,
which is moreover a complex integrable system in one of the complex
structures \cite{Donagi:1995cf}.  On the one hand, it was discovered
by Nekrasov and Shatashvili \cite{Nekrasov:2009rc} that an
$\Omega$-deformation on a two-plane quantizes a real symplectic
submanifold of the complex integrable system.  On the other hand, it
was found by Gaiotto, Moore and Neitzke \cite{Gaiotto:2010be} and Ito,
Okuda and Taki \cite{Ito:2011ea} that if the spacetime $\R^3 \times
S^1$ is replaced with a twisted product of $\R^3$ and $S^1$, then
supersymmetric loop operators form a noncommutative deformation of the
algebra of holomorphic functions on $\CM$ in other complex structures.

Despite the similarities between the two phenomena, explanations from
a unified point of view have been lacking.  In this paper we provide
such explanations, based on a connection that we establish between a
deformation of $\CN = 4$ supersymmetric sigma model and quantization
of symplectic manifolds.

More precisely, the main result of the paper concerns the
topologically twisted version of the sigma model, known as
Rozansky-Witten theory \cite{Rozansky:1996bq}.  We formulate it on a
three-manifold of the form $\R \times \Sigma$, with $\Sigma$ a Riemann
surface, taking the target space to be a hyperk\"ahler manifold $X$.
Given a complex structure on $X$ and a Killing vector field $V$ on
$\Sigma$, we construct a deformation of the theory analogous to the
$\Omega$-deformation in four dimensions.  In the case that $\Sigma$ is
a disk $D$, we show that the $\Omega$-deformed Rozansky-Witten theory
is equivalent to a quantum mechanical system whose phase space is a
symplectic submanifold of $X$ determined by the boundary condition.
The phenomena exhibited by the four-dimensional gauge theory then
follow as special cases of this result, applied to low-energy
effective sigma models with target space $X = \CM$.  The two cases
differ merely in the choice of complex structure.

While the four-dimensional phenomena are explained with a
three-dimensional theory, the construction of this theory is best
understood from a two-dimensional point of view.  To formulate the
$\Omega$-deformation for Rozansky-Witten theory on $\R \times \Sigma$
with target space $X$, we view the theory as a B-twisted
Landau-Ginzburg model \cite{Vafa:1990mu} on $\Sigma$ whose target
space is the space of maps from $\R$ to $X$.  For this reason we first
formulate the $\Omega$-deformation for general B-twisted
Landau-Ginzburg models, and then use this formulation to construct
the $\Omega$-deformed Rozansky-Witten theory.

The connection between the $\Omega$-deformed Rozansky-Witten theory
and quantization involves D-branes of novel type, which can be
introduced supersymmetrically in B-twisted Landau-Ginzburg models
only in the presence of the $\Omega$-deformation.  An amusing fact is
that in general these branes are similar to A-branes, rather than
B-branes.  There is an even more interesting analogy if we specialize
to the $\Omega$-deformed Rozansky-Witten theory.  In this case,
relevant branes are much like $(A, B, A)$-branes in $\CN = (4,4)$
supersymmetric sigma models: the support of a brane is the space of
maps from $\R$ to a submanifold $L$ of $X$ that is Lagrangian with
respect to the K\"ahler forms $\omega_I$ and $\omega_K$ associated to
two complex structures $I$ and $K$ of the hyperk\"ahler structure of
$X$, while holomorphic in the third complex structure $J$.  This
implies in particular that $L$ is a symplectic manifold with
symplectic form given by the restriction of $\omega_J$.

For $\Sigma = D$ with a brane of this type placed on the boundary, by
localization of the path integral we will derive a formula that
expresses a correlation function in terms of an integral over the
support of the brane.  In the context of Rozansky-Witten theory, the
localization formula gives an integration over maps from $\R$ to $L$.
This is nothing but the path integral for a quantum mechanical system
on $L$.  We therefore conclude that the $\Omega$-deformed
Rozansky-Witten theory quantizes the symplectic submanifold $(L,
\omega_J)$ of $X$.  The algebra of observables is found to be a
noncommutative deformation of the algebra of functions on $L$ that are
restrictions of holomorphic functions on $X$.

The appearance of $(A,B,A)$-like branes in our framework suggests a
relation to another approach to quantization, namely the one using
A-branes, developed by Gukov and Witten~\cite{Gukov:2008ve}.  Indeed,
the correspondence between $\Omega$-deformed $\CN = 2$ supersymmetric
gauge theories and quantum integrable systems was explained by
Nekrasov and Witten \cite{Nekrasov:2010ka} from this perspective.
(For explanations from other perspectives, see \cite{Bonelli:2011na,
  Aganagic:2011mi}.)  Their argument, however, relies on the fact that
the $\Omega$-deformation may be canceled away from the origin of the
two-plane by a redefinition of fields, and this makes the logic a
little involved.  Our approach hopefully renders the connection
between the $\Omega$-deformation and quantization more transparent.

In this paper we apply our framework to two specific problems in
four-dimensional gauge theory.  It will be interesting to find further
applications.  For example, the quantization of Seiberg-Witten curves
proposed in \cite{Fucito:2011pn} may find a natural place in the
present framework.  Besides, the framework itself may be generalized.
One direction in this regard would be to consider a gauged version of
Rozansky-Witten theory \cite{Kapustin:2009cd}, which is obtained by
topological twisting of an $\CN = 4$ supersymmetric gauge theory
constructed by Gaiotto and Witten \cite{Gaiotto:2008sd}.

Lastly, the $\Omega$-deformation of B-twisted theories should have a
broader range of applications.  The construction can be extended to
include gauge theories (the details of which will appear elsewhere),
and this extension may shed light on the correspondence between $\CN =
2$ superconformal theories in three dimensions and analytically
continued Chern-Simons theory~\cite{Dimofte:2010tz, Terashima:2011qi,
  Dimofte:2011jd, Dimofte:2011ju, Dimofte:2011py} via arguments along
the lines of \cite{Yagi:2013fda} (see also \cite{Lee:2013ida,
  Cordova:2013cea}).  Purely in two dimensions, it may prove fruitful
to study mirror symmetry between $\Omega$-deformed B-twisted theories
and $\Omega$-deformed A-twisted theories \cite{Shadchin:2006yz,
  Dimofte:2010tz}.

The rest of the paper is organized as follows.  In section~\ref{OBLG},
we formulate the $\Omega$-deformation of B-twisted Landau-Ginzburg
models and derive the localization formula.  In section~\ref{QORW}, we
construct the $\Omega$-deformed Rozansky-Witten theory and establish
its connection to quantization.  Section~\ref{4d} discusses the
applications to four-dimensional gauge theory.  In the appendix we
review the $\Omega$-deformation of twisted $\CN = 2$ supersymmetric
gauge theories in four dimensions.

\section{\texorpdfstring{$\boldsymbol \Omega$}{Omega}-deformation of B-twisted Landau-Ginzburg models}
\label{OBLG}

In this section we formulate the $\Omega$-deformation of B-twisted
Landau-Ginzburg models in two dimensions, based on which the
$\Omega$-deformed Rozansky-Witten theory is constructed.
Furthermore, we study boundary conditions in the presence of the
$\Omega$-deformation, and derive a localization formula for
correlation functions on a disk.  The results obtained here will be
essential in our discussion in the next section.

\subsection[\texorpdfstring{$\Omega$}{Omega}-deformed B-twisted Landau-Ginzburg models]{\texorpdfstring{$\boldsymbol \Omega$}{Omega}-deformed B-twisted Landau-Ginzburg models}

Let us recall how the $\Omega$-deformation in four dimensions works
\cite{Nekrasov:2002qd, Nekrasov:2003rj}.  A topologically twisted $\CN
= 2$ supersymmetric gauge theory \cite{Witten:1988ze} has a single
scalar supercharge $Q$, satisfying the relation $Q^2 = 0$ modulo a
gauge transformation (as well as a flavor symmetry transformation if
hypermultiplet masses are nonzero \cite{Hyun:1995hz,
  Labastida:1996tz}).  This is used as a BRST operator, meaning that
physical operators and states are $Q$-cohomology classes.  To
introduce an $\Omega$-deformation, one chooses a vector field $V$
generating an isometry of the spacetime four-manifold with respect to
a given metric.  With this choice understood, the BRST operator of the
$\Omega$-deformed theory obeys the deformed relation
\begin{equation}
  \label{Q2LV}
  Q^2 = L_V,
\end{equation}
where $L_V$ is the conserved charge that acts on fields as the Lie
derivative $\CL_V$ by $V$.  The construction of the $\Omega$-deformed
theory is reviewed in the appendix.

Similarly, a B-twisted Landau-Ginzburg model in two dimensions
\cite{Vafa:1990mu} has a BRST operator $Q$ satisfying $Q^2 = 0$ up to
a central charge.  In order to formulate an $\Omega$-deformation of
this theory, we should therefore pick a Killing vector field $V$ on
the worldsheet $\Sigma$ equipped with a hermitian metric $h$, and
deform the theory so that the modified BRST operator obeys the
deformed relation \eqref{Q2LV}.  This is what we are aiming for.%
\footnote{In \cite{Closset:2014pda}, a supergravity background was
  found that realizes the $\Omega$-deformation of A-twisted theories
  on $S^2$.  As mentioned in that paper, one can combine it with a
  $\Z_2$-action implementing mirror symmetry to obtain an
  $\Omega$-deformation of B-twisted theories.  Our formulation
  presumably reproduces their results for $\Sigma = S^2$.  I would
  like to thank Stefano Cremonesi for explaining their work.}

How can we achieve such a deformation?  To get the idea, consider the
simplest case $\Sigma = \C$.  In this case the theory retains the full
$\CN = (2,2)$ supersymmetry in the twisted form, generated by two
scalar supercharges $\Qb_+$, $\Qb_-$ and a one-form supercharge $G =
G_z \rmd z + G_\zb \rmd\zb$.  These satisfy $\{\Qb_-, G_z\} = P_z$ and
$\{\Qb_+, G_\zb\} = P_\zb$, where $P = P_z \rmd z + P_\zb \rmd\zb$ is
the generator of translations; the other commutators either vanish or
give central charges.  The BRST operator of the undeformed theory is
$Q = \Qb_+ + \Qb_-$.  If we take a vector field $V = V^z \del_z +
V^\zb\del_\zb$ with constant components $V^z$, $V^\zb$ and modify the
BRST operator to $Q = \Qb_+ + \Qb_- + \iota_V G$, then we obtain the
desired relation $Q^2 = \iota_V P$.  We are going to generalize this
construction to an arbitrary worldsheet $\Sigma$.

The target space of the theory is a K\"ahler manifold $Y$.
Classically the following construction makes sense for any K\"ahler
target space, but for quantum anomalies to be absent, $Y$ has to be
Calabi-Yau.  (This is essentially due to the fact that the axial
$\U(1)$ R-symmetry used in the B-twist is anomalous unless $c_1(Y) =
0$.)  We denote the holomorphic and antiholomorphic tangent bundles of
$Y$ by $TY$ and $\TYb$, respectively, and their duals with superscript
$\vee$.  We pick a K\"ahler metric $g$ on $Y$.

The bosonic field of the theory is a map $\Phi\colon \Sigma \to Y$.
Given local coordinates on $Y$, we can express $\Phi$ locally by a set
of functions $(\phi^i, \phib^\ib)$.  In the standard formulation, the
fermionic fields of the B-twisted theory are scalars $\eta$ with
values in $\TYb$ and $\theta$ with values in $T^\vee Y$, and a
one-form $\rho$ with values in $TY$.  In our construction, we use
instead of $\theta$ a two-form $\mu$ with values in $\TYb$; the two
are related by the Hodge duality and the isomorphism between $\TYb$
and $T^\vee Y$ induced by $g$.  Thus the fermionic fields of the
theory are
\begin{equation}
  \eta \in \Omega^0(\Sigma, \Phi^*\TYb), \quad
  \rho \in \Omega^1(\Sigma, \Phi^*TY), \quad
  \mu \in \Omega^2(\Sigma, \Phi^*\TYb).
\end{equation}
We also introduce auxiliary bosonic fields.  They are two-forms $F$
with values in $TY$ and $\Fb$ with values in $\TYb$:
\begin{equation}
  F \in \Omega^2(\Sigma, \Phi^*TY), \quad
  \Fb \in \Omega^2(\Sigma, \Phi^*\TYb).
\end{equation} 

Starting from $\CN = (2,2)$ supersymmetry transformation laws for
B-twisted chiral multiplets \cite{Labastida:1991qq}, it is
straightforward to write down the $\Omega$-deformed supersymmetry
transformation laws, following the same procedure as in the flat case
described above.  After shifting $\Fb$ to absorb dependence on the
worldsheet metric, we have
\begin{equation}
  \label{SUSY}
  \begin{alignedat}{2}
    \delta\phi^i &= \iota_V\rho^i, &
    \delta\phib^\ib &= \eta^\ib,
    \\
    \delta_\nabla\rho^i
    &= \rmd\phi^i + \iota_V F^i, &
    \delta_\nabla\eta^\ib &= V(\phib^\ib),
    \\
    \delta_\nabla F^i
    &= \rmd_\nabla\rho^i
       + \frac12 R^i{}_{j\kb l}\eta^\kb \rho^j \wedge \rho^l, &
    \delta_\nabla\mu^\ib &= \Fb^\ib,
    \\
    & & \quad
    \delta_\nabla\Fb^\ib
    &= \rmd_\nabla \iota_V\mu^\ib
       + R^\ib{}_{\jb k\lb} \iota_V\rho^k \eta^\lb \mu^\jb.
  \end{alignedat}
\end{equation}
Here $\delta_\nabla$ is the supersymmetry variation coupled to the
pullback of the Levi-Civita connection $\nabla$ of $g$; for example,
$\delta_\nabla \rho^i = \delta\rho^i + \delta\phi^k \Gamma^i_{kj}
\rho^j$.  (More generally, $\nabla$ can be any torsion-free connection
on $TY$ whose curvature form $R$ is of type $(1,1)$ and obeys the
first Bianchi identity.)  Notice the similarity to the supersymmetry
transformation laws \eqref{SUSY-4d} for $\Omega$-deformed theories in
four dimensions.

One can verify that the ``raw'' supersymmetry variation $\delta$
satisfies $\delta^2 = \CL_V$.%
\footnote{An easy way to see this is to define $G^i = F^i - \frac12
  \Gamma^i_{kj} \rho^k \wedge \rho^j$ and $\Gb^\ib = \Fb^\ib -
  \Gamma^\ib_{\kb\jb} \eta^\kb\mu^\jb$, in terms of which one has
  $\delta\rho^i = \rmd\phi^i + \iota_V G^i$, \ $\delta G^i =
  \rmd\rho^i$ and $\delta\mu^\ib = \Gb^\ib$, \ $\delta\Gb^\ib =
  \rmd\iota_V\mu^\ib$.}
As usual, we let $Q$ denote the generator of the supersymmetry
variation $\delta$.  Then it obeys the deformed relation \eqref{Q2LV},
as desired. Strictly speaking, on the right-hand side of this relation
may appear an extra conserved charge that commutes with fields and
coincides for $V = 0$ with a central charge of the B-twisted $\CN = 2$
supersymmetry algebra.  Although such an extra term is important when
one considers the action of $Q$ on states, it plays no role in our
discussion and hence will be ignored.

For the moment we assume that $\Sigma$ has no boundary; we will
discuss boundary effects shortly.  Then the $Q$-invariant action $S$
of the $\Omega$-deformed theory consists of two pieces, $S = S_0 +
S_W$.  The first piece $S_0$ is $Q$-exact and contains kinetic terms.
The second piece $S_W$ is constructed from a superpotential $W\colon Y
\to \C$, which is a holomorphic function of $\phi^i$.  Unlike $S_0$,
this one is not $Q$-exact.

Concretely, we can take $S_0$ to be
\begin{equation}
  \label{S0}
  S_0
  = \delta\int_\Sigma \Bigl(g_{i\jb}\rho^i
    \wedge \star\bigl(\rmd\phib^\jb + \iota_V\Fb^\jb\bigr)
        + g_{i\jb} F^i \wedge \star\mu^\jb\Bigr),
\end{equation}
while $S_W$ is given by
\begin{equation}
  \label{SW}
  S_W
  = i\int_\Sigma\Bigl(
    F^i \del_i W
    + \frac{1}{2} \rho^i \wedge \rho^j \nabla_i\del_j W
    + \delta\bigl(\mu^\ib \del_\ib\Wb\bigr)
    \Bigr).
\end{equation}
The $Q$-invariance of the action relies on the assumption that $V$ is
a Killing vector field.  This ensures that $\CL_V$ commutes with
$\star$, so if we write the $Q$-exact part of the Lagrangian as
$\delta\CV$, then $\delta^2\CV = \CL_V\CV = \rmd\iota_V\CV$ by the
formula $\CL_V = \rmd\iota_V + \iota_V\rmd$.

Computing the supersymmetry variation we find
\begin{multline}
  S_0
  = \int_\Sigma \Bigl(
  g_{i\jb} \bigl(\rmd\phi^i + \iota_VF^i\bigr)
  \wedge \star\bigl(\rmd\phib^\jb + \iota_V\Fb^\jb\bigr)
  + g_{i\jb} F^i \wedge \star\Fb^\jb
  - g_{i\jb} \rho^i \wedge \star\rmd_\nabla\eta^\jb
  + g_{i\jb} \rmd_\nabla\rho^i \wedge \star\mu^\jb \\
  + \frac12 R_{\ib j\kb l} \eta^\kb \rho^j \wedge \rho^l \wedge \star\mu^\ib
  - \rho^i
    \wedge \star\iota_V\bigl(g_{i\jb} \rmd_\nabla \iota_V\mu^\jb
    + R_{i\jb k\lb} \iota_V\rho^k \eta^\lb \mu^\jb\bigr)
    \Bigr).
\end{multline}
For the superpotential terms we get
\begin{equation}
  S_W
  = i\int_\Sigma\Bigl(
    F^i \del_i W
    + \frac{1}{2} \rho^i \wedge \rho^j \nabla_i\del_j W
    + \Fb^\ib \del_\ib\Wb
    + \eta^\ib\mu^\jb\nabla_\ib\del_\jb\Wb
    \Bigr).
\end{equation}
Integrating out the auxiliary fields produces the potential
\begin{equation}
  \label{F-term}
  \star\|\rmd W\|^2 + \dotsb
  = \star(g^{i\jb} \del_i W \del_\jb\Wb) + \dotsb,
\end{equation}
where we have abbreviated the terms involving $V$.
%
%
One can add more $Q$-exact terms to the Lagrangian if one wishes, such
as $\delta(g_{\ib j} \eta^\ib V(\phi^j))$ which produces the
$V$-dependent potential $\|V(\phi)\|^2$.

When $V = 0$, the supersymmetry transformation and the action
constructed above reduce to those of the ordinary B-twisted
Landau-Ginzburg model, up to the replacement of $\theta$ with $\mu$
explained above.  This construction therefore defines a deformation of
the latter theory for $V \neq 0$.

The worldsheet metric $h$ appears in the action only through the Hodge
duality within the $Q$-exact piece $S_0$.  Hence, the
$\Omega$-deformed theory is quasi-topological, in the sense that it is
invariant under deformations of the metric as long as $V$ remains to
be a Killing vector field.  In addition, the theory is invariant under
overall rescaling of the target space metric $g$, as this leaves the
supersymmetry transformation invariant and the metric enters the
action through $S_0$.

The observables of the theory are the $Q$-closed operators that are
not $Q$-exact.  At the zeros of $V$, any local observables of the
undeformed theory remain to be observables.  There is a class of local
observables that are in one-to-one correspondence with the elements of
the $\delb$-cohomology $H^{0,\bullet}(Y; \C)$.  To see this, note that
for $V = 0$, the action of $Q$ on $\phi^i$, $\phib^\ib$ and $\eta$
coincides with that of $\delb$ under the identification of $\eta^\ib$
with $\rmd\phib^\ib$.  Thus, if $\omega = \omega_{\ib_1 \dotso \ib_q}
\rmd\phib^{\ib_1} \wedge \dotsm \wedge \rmd\phib^{\ib_q}$ is a
$\delb$-closed $(0,q)$-form on $Y$, then $\omega_{\ib_1 \dotso \ib_q}
\eta^{\ib_1} \dotsm \eta^{\ib_q}$ inserted at a zero of $V$ is a
$Q$-closed local operator, and represents a nonzero $Q$-cohomology
class if and only if $\omega$ represents a nonzero element in
$H^{0,\bullet}(Y;\C)$.  In particular, holomorphic functions on $Y$
correspond to local observables.

\subsection{Incorporating boundaries}

So far we have assumed that $\Sigma$ has no boundary.  Now we consider
the situation that $\Sigma$ has a boundary and the isometry generated
by $V$ restricts to an isometry of $\del\Sigma$.  In this situation
the $Q$-invariance of the action must be reexamined.  Also, we have to
ask what sort of boundary conditions are physically sensible.  In the
following we will express the $Q$-variation by $Q$-commutator,
reserving $\delta$ for arbitrary variation of fields in order to avoid
possible confusion.

Let us address the issue of $Q$-invariance.  The $Q$-exact part of the
action remains to be $Q$-invariant in the presence of boundary.  For,
if $\CV$ is a two-form, then
\begin{equation}
  \int_\Sigma [Q, \{Q, \CV\}]
  = \int_\Sigma (\rmd\iota_V + \iota_V\rmd)\CV
  = \int_{\del\Sigma} \iota_V \CV
  = 0.
\end{equation}
The last equality follows from the assumption that $V$ generates an
isometry of $\del\Sigma$ and hence is tangent to $\del\Sigma$.  The
potential problem therefore comes from the non-$Q$-exact part.
Indeed, its $Q$-variation gives
\begin{equation}
  i\int_{\del\Sigma} \rho^i \del_i W,
\end{equation}
breaking the $Q$-invariance by a boundary contribution.

We must somehow eliminate this boundary contribution to recover the
$Q$-invariance.  In the case of ordinary B-twisted Landau-Ginzburg
models, one can do this by imposing a B-brane boundary condition.
This condition requires that $\Phi$ map $\del\Sigma$ to a
submanifold~$\gamma$ of $Y$,
\begin{equation}
  \Phi(\del\Sigma) \subset \gamma,
\end{equation}
and $\rmd W|_\gamma = 0$, that is, $W$ be locally constant on
$\gamma$.  The boundary contribution vanishes then.

In the $\Omega$-deformed case, there is another way of eliminating the
boundary contribution.  For simplicity, suppose that there is only one
connected boundary component and it is compact.  Let $\vphi$ be a
periodic coordinate on $\del\Sigma$ such that $h_{\vphi\vphi}$ is
constant.  In this coordinate,
\begin{equation}
  V|_{\del\Sigma} = \veps\del_\vphi
\end{equation}
for some real constant $\veps$.  Assuming that $\veps \neq 0$, we can
add to the action the boundary term
\begin{equation}
  \label{BT}
  -\frac{i}{\veps} \int_{\del\Sigma} \rmd\vphi \, (W + W_0),
\end{equation}
where $W_0$ is a locally constant function on $\gamma$.  The
$Q$-variation of this term cancels the boundary contribution in
question, recovering the $Q$-invariance of the action.

One interpretation of the above boundary term is that it is the action
for a theory living on the boundary, with $\veps$ being the Planck
constant.  The undeformed limit $\veps \to 0$ is the classical limit,
and in this limit $\Phi$ obeys the equation of motion $\rmd W = 0$ on
the boundary, which reproduces the ordinary B-brane condition on $W$.

This mechanism of recovering the $Q$-invariance is interesting since
it is available only when the $\Omega$-deformation is turned on.
Moreover, it requires a weaker boundary condition on $W$ compared to
the B-brane condition.  For the boundary term \eqref{BT} to not spoil
the convergence of the path integral, its real part had better be
nonnegative.  For our purposes it is sufficient to consider the
situation that the boundary term is purely imaginary.  To ensure this
property, we place on the boundary a brane supported on $\gamma$, and
impose
\begin{equation}
  \label{ImdW=0}
  \Im\rmd W|_\gamma = 0.
\end{equation}
Then, the constant imaginary part of $W$ can be absorbed into $W_0$,
and the boundary term can be written as
\begin{equation}
  \label{BT-2}
  -\frac{i}{\veps} \int_{\del\Sigma} \rmd\vphi \, (\Re W + W_0),
\end{equation}
with $W_0$ now chosen to be real.

We would like to write down a set of boundary conditions that defines
this brane.  A guiding principle for determining physically sensible
conditions is that in a weak coupling limit, solutions to equations of
motion should be saddle point configurations of the path integral.  In
other words, when the fields are varied in that limit, boundary terms
should not arise in the variation of the action.  Recalling that in
our case the bulk theory is invariant under rescaling of the target
space metric $g$, we see that there is a natural weak coupling limit,
namely the limit in which $g$ is rescaled by a large factor.  This is
also the limit we will consider in the derivation of the localization
formula for correlation functions.%
\footnote{Actually we will consider a slightly different limit which
  simplifies the analysis, but the two limits lead to the same
  boundary condition.}
 We therefore define our brane as the boundary condition
obtained by taking variations in this limit.  (A similar choice was
made in \cite{Hori:2013ika} where $\CN = (2,2)$ supersymmetric
theories on a hemisphere were studied.)

Since $S_0$ dominates in the limit under consideration, we can ignore
the remaining part of the action in our analysis.  Setting $\delta S_0
= 0$ and using the equations of motion for $F$ and $\Fb$ derived from
$S_0$, we find the constraints
\begin{equation}
  g(\delta\Phi, \iota_V\star\rmd\Phi)|_{\del\Sigma}
  =  g\bigl([Q,\delta\Phi], (\iota_V\star\rho, \star\mu)\bigr)|_{\del\Sigma}
  = 0.
\end{equation}
The boundary condition for $\Phi$ implies that any variation of $\Phi$
is tangent to $\gamma$ on $\del\Sigma$, and we require that the same
be true for the $Q$-variation $[Q,\Phi]$; thus we have
\begin{equation}
  \label{QPhi-in-Tgamma}
  (\iota_V\rho, \eta) \in T_\C\gamma
\end{equation}
at each point on $\del\Sigma$.  Assuming that the variation of $\Phi$
is not constrained in any other way, we conclude that
\begin{equation}
  \iota_V\star\rmd\Phi \in N_\R\gamma, \quad
  (\iota_V\star\rho, \star\mu) \in N_\C\gamma,
\end{equation}
where $N_\R\gamma$ is the normal bundle of $\gamma$ and $N_\C\gamma$
is its complexification.  In particular, $\Phi$ obeys the Neumann
boundary condition in the direction normal to the boundary, just as in
the case of ordinary B-branes (with vanishing $B$-field and Chan-Paton
gauge field).

Furthermore, in order for $Q$ to act on the space of allowed field
configurations, the boundary condition itself must be invariant under
the action of $Q$ (or more precisely, the covariant version of it,
coupled to the Levi-Civita connection).  This leads to additional
constraints generated by repeated action of $Q$ on the constraints
described above.  An example is the constraint \eqref{QPhi-in-Tgamma},
which comes from the D-brane constraint $\Phi(\del\Sigma) \subset
\gamma$.  This procedure generates only a few new constraints since
$Q^2 = L_V$ leaves invariant the space of sections of a vector bundle
over $\del\Sigma$.  It turns out that these additional constraints
follow from the constraints discussed already if we use the equations
of motion derived from the quadratic part of $S_0$.

The above boundary condition is independent of $g$, thanks to the
limit considered here which decouples dependence on the
superpotential.  It is also independent of the component of the
worldsheet metric $h$ normal to $\del\Sigma$.  (If $n$ is a coordinate
in the normal direction, we have $\iota_V\star\rmd\Phi |_{\del\Sigma}
= \veps \sqrt{h_{\vphi\vphi}/h_{nn}} (\del_n\phi^i, \del_n\phib^\ib)$
and $(\iota_V\star\rho, \star\mu) |_{\del\Sigma} =
\sqrt{h_{\vphi\vphi}/h_{nn}} (\veps \rho^i_n, h^{\vphi\vphi}
\mu^\ib_{n\vphi})$.)  Hence, the invariance of the bulk theory under
relevant deformations of the metrics is mostly preserved by the brane,
broken only by the explicit dependence on $h_{\vphi\vphi}$.

\subsection{Localization}

As in ordinary A- and B-twisted theories, the path integral for a
correlation function of $Q$-invariant operators in the
$\Omega$-deformed B-twisted Landau-Ginzburg model reduces to an
integral over a small subspace of the field space.  Let us derive a
formula for correlation functions in the case that $\Sigma$ is a disk
$D$, with a brane of the above type placed on the boundary.

We equip $D$ with the metric of the form $h = h_{rr}(r) \rmd r^2 +
h_{\vphi\vphi}(r)\rmd\vphi^2$, where $(r, \varphi)$ are polar
coordinates.  Then $V = \veps \del_\varphi$, with $\veps$ constant.
Here $\veps$ is real, but it is also possible to make it complex since
$V$ only needs to satisfy the Killing equation.

Our theory is invariant under rescaling of the target space metric
$g$.  In particular, we can rescale it as $g \to t^2 g$ and take the
limit $t \to \infty$.  Integrating out the auxiliary fields, we find
that in this limit the action diverges away from the locus where
\begin{equation}
  \rmd\phi^i = 0.
\end{equation}
The path integral therefore localizes to the constant maps, that is to
say, receives contributions solely from an arbitrarily small
neighborhood of the subspace of constant maps in the space of maps
from $D$ to $Y$.

Such a neighborhood may be thought of as a fibration over the space of
constant maps.  By the boundary condition, the constant maps are
required to map into $\gamma \subset Y$, the support of the brane.
Thus the base is isomorphic to $\gamma$.  The fiber consists of the
bosonic fluctuation $\vphi$ around a constant map.  We can extend the
fiber so that it includes the fermionic fields as well.  The path
integral is then an integral over the total space of the extended
fibration.  What we want to do now is to perform the integration over
the fiber, and reduce the path integral to an integral over the base.

It may be helpful to recall how the fiber integration is done in a
simpler setting where $V = 0$ and $\Sigma$ has no boundary.  We
combine $\eta$ and $\mu$ into a single field $\zeta = -\eta + \mu$
which is an even-degree form with values in $\TYb$.  To quadratic
order, the part of the action relevant for large $t$ can be written as
\begin{equation}
  \label{S-quad}
  t^2\bigl(\langle\vphi, \Delta \vphi\rangle
  + \langle (\rmd_\nabla + \rmd_\nabla^*)\rho, \zeta\rangle
  + \langle\rho, (\rmd_\nabla + \rmd_\nabla^*)\zeta\rangle\bigr).
\end{equation}
Here $\Delta = (\rmd_\nabla + \rmd_\nabla^*)^2$, and $\langle\cdot,
\cdot\rangle$ is an inner product defined by the metric on $\Sigma$
and the original metric on $Y$ before the rescaling.  We expand
$\vphi$, $\rho$ and $\zeta$ in orthonormal bases of eigenmodes of
$\Delta$, and express the quadratic part of the action in terms of the
expansion coefficients.  The integration over the fiber is integration
over these coefficients.  We can rescale the coefficients by $1/t$ to
absorb the overall $t^2$ factor in the quadratic part.  Provided that
there are no fermion zero modes, after doing so the terms of higher
order are suppressed by inverse powers of $t$.  In the limit $t \to
\infty$, the fiber integration produces the ratio of the bosonic and
fermionic one-loop determinants, $\prod_\beta
\sqrt{\lambda'_\beta}/\prod_\alpha \lambda_\alpha$, where
$\lambda_\alpha$ are nonzero eigenvalues for $\vphi$, and
$\lambda'_\beta$ are those for $\rho$ and $\zeta$.  The nonzero
eigenvalues for $\rho$ and $\zeta$ agree since their nonzero modes are
related by the action of $\rmd_\nabla + \rmd_\nabla^*$, which commutes
with $\Delta$.  Similarly, as $\Delta$ and $\star$ commute, the
nonzero modes for zero- and two-forms are related by the Hodge duality
and have the same eigenvalues.  This means that the set
$\{\lambda'_\beta\}$ consists of two copies of the set
$\{\lambda_\alpha\}$.  Hence, this ratio is equal to $1$, and the
fiber integration is trivial in this case.

We want to carry out a similar computation in the case at hand, where
$V$ generates rotations of $\Sigma = D$.  Here the analysis is a bit
more complicated.

One complication is that if $V \neq 0$, the action contains additional
terms and they modify the quadratic part.  To simplify the analysis,
we replace the $Q$-exact piece $S_0$ of the action with
\begin{equation}
  t^2 \delta\int_\Sigma \Bigl(g_{i\jb}\rho^i
  \wedge \star\bigl(\rmd\phib^\jb + \iota_V\Fb^\jb\bigr)
  + s g_{i\jb} F^i \wedge \star\mu^\jb\Bigr),
\end{equation}
rescale $\mu \to \mu/s$, and take the limit $s \to \infty$.
Integrating out the auxiliary fields and performing integration by
parts using the boundary condition, we find that the relevant part of
the on-shell action now takes the identical form \eqref{S-quad} as in
the case with $V = 0$.

Another complication comes from the presence of boundary, which makes
the analysis of mode expansion more difficult.  We can deal with this
problem as follows.  Using the freedom of deforming the worldsheet
metric, we can choose $D$ to have the shape of a sphere $S^2$ with a
small hole; in spherical coordinates $(\vtheta, \vphi)$, the metric
takes the form $h = R^2(\rmd\vtheta^2 + \sin^2\vtheta\rmd\vphi^2)$,
with $\vtheta$ ranging from $0$ to some value $\vtheta_{\del D}$ where
the boundary is located.  (Our boundary condition depends on
$h_{\vphi\vphi}$.  It should be given with respect to the original
metric and fixed throughout the deformation.)  Then we take the limit
$\vtheta_{\del D} \to \pi$.  In this limit $D$ becomes the whole
$S^2$, and the boundary state gets mapped to a $Q$-invariant local
operator inserted at $\vtheta = \pi$.  If we now expand the fields in
the eigenmodes of $\Delta$ on $S^2$, then in terms of the expansion
coefficients the quadratic part of the action has the same expression
as before.

The conclusion is therefore that the fiber integration is again exact
at one loop and produces a factor similar to the ratio of the bosonic
and fermionic determinants, assuming that there are no fermion zero
modes.  The difference is that this time the path integral receives
contributions only from the locus where the expansion coefficients
obey various relations, imposed by the boundary state or the operator
inserted at $\vartheta = \pi$.

The question is whether this one-loop factor depends on the background
constant map $\Phi_0$ around which we are expanding.  If it does, the
dependence should come from $g$, $W$ or $\gamma$, since these are the
only objects defined on the target space that enter our setup.  The
one-loop computation refers to just the quadratic part of the action
in the limit $t \to \infty$, and this is independent of $W$.  It is
also independent of $g$ if we use holomorphic normal coordinates
centered at $\Phi_0$, in which $g_{i\jb}(\Phi_0) = \delta_{ij}$ and
$\del_k g_{i\jb}(\Phi_0) = \del_\kb g_{i\jb}(\Phi_0) = 0$.  In fact,
in these coordinates the quadratic part takes exactly the same form as
the action for an affine target space $\C^n$ with the standard metric.

This leaves $\gamma$ as the only possible source for nontrivial
dependence on the background.  Indeed, the choice of $\gamma$ may
introduce such dependence, since it specifies the boundary condition
which in turn determines the relations among the mode expansion
coefficients.  Put another way, the one-loop factor is independent of
the background if we can choose $\gamma$ in such a way that the
boundary condition is the same for all backgrounds.  In view of the
fact that the boundary condition in a background $\Phi_0 \in \gamma$
is determined by the tangent and normal spaces of $\gamma$ at
$\Phi_0$, a sufficient condition for background independence is that
the tangent spaces (and hence also the normal spaces) at any two
points of $\gamma$ can be made identical, when regarded as subspaces
of $\C^n$ via a suitable choice of normal coordinate systems centered
at these points.  Taking into account the freedom in choosing normal
coordinates, we see that the two spaces need to be identical up to an
action of $\U(n)$.

One way to satisfy this condition is to take $\gamma$ to be a complex
submanifold of $Y$, as in the case for ordinary B-branes.  In this
case the superpotential $W$ restricts to a holomorphic function on
$\gamma$.  As we require $\Im\rmd W|_\gamma = 0$, $W$ would then have
to be locally constant on $\gamma$.  This is in fact the ordinary
B-brane condition on $W$.  However, it is not desirable for our
purposes.  We would like to view $W$ as the Lagrangian of a boundary
theory, from which the equation $\rmd W = 0$ follows as a classical
equation of motion.

A more interesting possibility is to take $\gamma$ to be a Lagrangian
submanifold of $Y$ with respect to the K\"ahler form.  The condition
for background independence is then satisfied since $\U(n)$ acts
transitively on the Lagrangian Grassmannian $\U(n)/\OO(n)$, the space
of Lagrangian subspaces of $\R^{2n}$.  From now on we will consider
this kind of supports.

Finally, we have to make sure that the assumption of absence of
fermion zero modes is actually true.  Since the result of the path
integral is independent of the size of the $S^2$, we will show this in
the limit where the $S^2$ is very small.  First of all, there are no
harmonic one-forms on $S^2$ and hence no zero modes for $\rho$.  The
zero modes of $\eta$ are constants, while those of $\mu$ are their
Hodge duals. For these modes, we have to look at the constraints
imposed by the boundary condition.  In the limit we are considering,
the nonzero modes are very massive and decouple, so the fermions can
be replaced by their zero mode parts.  Then the boundary condition
forces $(0, \eta) \in T_\C \gamma$ and $(0, \star\mu) \in N_\C
\gamma$.  Since $\gamma$ is a Lagrangian submanifold of a K\"ahler
manifold, we have $I(T_\R \gamma) = N_\R \gamma$ and it follows that
$\eta = \mu = 0$ on the boundary.  Hence, the zero modes of $\eta$ and
$\mu$ are actually identically zero in this limit.

We have found that the fiber integration just produces an irrelevant
constant.  The remaining step in the path integral is to integrate
over all background constant maps.  For a constant map $\Phi_0$, the
action is evaluated as
\begin{equation}
  S(\Phi_0) = -\frac{2\pi i}{\veps} \bigl(\Re W + W_0\bigr)(\Phi_0).
\end{equation}
Altogether, we conclude that the path integral for the
$\Omega$-deformed B-twisted Landau-Ginzburg model on a disk reduces
to the form
\begin{equation}
  \label{LF}
  \vev{\CO}
  = \int_\gamma \rmd\Phi_0
    \exp\Bigl(\frac{2\pi i}{\veps} \bigl(\Re W + W_0\bigr)(\Phi_0)\Bigr)
    \CO(\Phi_0),
\end{equation}
where the operator insertion on the right-hand side is evaluated for
constant maps $\Phi_0 \in \gamma$, with fermions set to zero.  In this
expression we have renormalized $W_0$ to absorb the one-loop factor.

\section{Quantization via the
  \texorpdfstring{$\boldsymbol\Omega$}{Omega}-deformed Rozansky-Witten theory}
\label{QORW}

Having constructed $\Omega$-deformed B-twisted Landau-Ginzburg
models, we now move up one dimension higher and formulate the
$\Omega$-deformation of Rozansky-Witten theory in three dimensions.
We will then establish, via localization of the path integral, the
connection between the $\Omega$-deformed Rozansky-Witten theory and
quantization of symplectic submanifolds of the hyperk\"ahler target
space.

\subsection{Rozansky-Witten theory}

To begin, let us review the basic aspects of Rozansky-Witten theory.
We refer the reader to the original paper \cite{Rozansky:1996bq} for
more details.

Rozansky-Witten theory is a three-dimensional supersymmetric sigma
model which can be defined on a general three-manifold $M$.  The
target space of the theory is a complex symplectic manifold $(X,
\Omega)$.  It is a complex manifold $X$ equipped with a nondegenerate
closed holomorphic two-form $\Omega$, called a holomorphic symplectic
form.

Let $\Phi\colon M \to X$ be the bosonic map of the sigma model, and
write $(\phi^i, \phib^\ib)$ for a local expression of $\Phi$.  The
theory has two fermionic fields, $\eta$ and $\chi$, and is invariant
under the following supersymmetry:
\begin{equation}
  \label{Q-RW}
  \begin{alignedat}{2}
    \delta\phi^i &= 0, &\qquad
    \delta\phib^\ib &= \eta^\ib,
    \\
    \delta\chi^i &= \rmd\phi^i, &
    \delta\eta^\ib &= 0.
  \end{alignedat}
\end{equation}
As can be seen from the transformation laws, $\eta$ is a scalar on $M$
with values in $\overline{TX}$, and $\chi$ is a one-form on $M$ with
values in $TX$.  The generator $Q$ of the supersymmetry satisfies $Q^2
= 0$.  We use it as a BRST operator, declaring that physical operators
and states are $Q$-cohomology classes.

To construct a $Q$-invariant action we need to make some choices.  We
pick a Riemannian metric $h$ on $M$ and a hermitian metric $g$ on $X$.
In addition, we choose a torsion-free connection $\nabla = \rmd +
\Gamma$ on $TX$, with connection matrices $\Gamma^i_{kj} =
\Gamma^i_{jk}$.  The $(1,1)$-part of the curvature form of $\nabla$
(given by the matrix elements $R^i{}_{jk\lb} \rmd\phi^k \wedge
\rmd\phib^\lb$ with $R^i{}_{j\lb k} = \del_\lb\Gamma^i_{kj}$)
represents a $\delb$-cohomology class, known as the Atiyah class of
$X$.  It is the obstruction to the existence of a holomorphic
connection on $TX$.

The action is the sum of two pieces, $S = S_1 + S_2$.  The first piece
is $Q$-exact:
\begin{equation}
  \label{S1}
  S_1 = \delta\int_M
        g_{i\jb} \chi^i \wedge \star\rmd\phib^\jb
      = \int_M\bigl(g_{i\jb} \rmd\phi^i \wedge \star\rmd\phib^\jb
          - g_{i\jb} \chi^i \wedge \star\rmd_{\nablat}\eta^\jb\bigr).
\end{equation}
Here the connection $\nablat = \rmd + \Gammat$ is defined by
$(\Gammat_\kb)^\ib{}_j = g^{\ib l} \del_j g_{l\kb}$; if $g$ is
K\"ahler, $\nablat$ is the Levi-Civita connection of $g$.  The second
piece is $Q$-invariant, but not $Q$-exact:
\begin{equation}
  \label{S2}
  S_2
  = -\frac{i}{4} \int_M\Bigl(
    \Omega_{ij} \chi^i \wedge \rmd_\nabla\chi^j
    - \frac{1}{3} \Omega_{ij} R^j{}_{kl\mb}
      \chi^i \wedge \chi^k \wedge \chi^l \eta^\mb
    + \frac{1}{3} \nabla_k \Omega_{ij}
      \rmd\phi^i \wedge \chi^j \wedge \chi^k\Bigr).
\end{equation}
The particular normalization is chosen for later convenience.

Since the spacetime metric $h$ appears only in the $Q$-exact part
$S_1$ through the Hodge duality, the theory is topological.  Likewise,
the target space metric $g$ appears only in $S_1$, so the theory is
independent of the choice of $g$.  It turns out that the theory is
also independent of the choice of the connection $\nabla$, different
choices leading to the same expression for $S_2$ modulo $Q$-exact
terms.

The local observables of Rozansky-Witten theory are in one-to-one
correspondence with the $\delb$-cohomology classes of $X$ under the
identification of $\eta^\ib$ with $\rmd\phib^\ib$.  There are also
nonlocal observables.  Given a connection $A$ of type $(1,0)$ on any
holomorphic $G$-bundle $E \to X$, we define a $Q$-invariant connection
\begin{equation}
  \CA = A_i \rmd\phi^i + F_{i\jb} \chi^i\eta^\jb,
\end{equation}
where $F$ is the curvature of $A$.  Using this connection we can
construct a $Q$-invariant loop operator
\begin{equation}
  \Tr P\exp\Bigl(\oint_\CC \CA\Bigr),
\end{equation}
by taking the trace of the holonomy of $\CA$ along a closed path $\CC$
in $M$.

A special case of interest is when $X$ admits a hyperk\"ahler
structure $(g, I, J, K)$.  In this case, $X$ has a two-sphere $\P^1$
of complex structures, and $g$ is K\"ahler with respect to all of
them.  Elements of the $\P^1$ are linear combinations $aI + bJ + cK$,
with $a^2 + b^2 + c^2 = 1$ and $I$, $J$, $K$ satisfying the quaternion
relations
\begin{equation}
  I^2 = J^2 = K^2 = IJK = -1.
\end{equation}
If we write $\omega_J$ and $\omega_K$ for the K\"ahler forms
associated to $J$ and $K$, respectively, then
\begin{equation}
  \label{Omega_I}
  \Omega_I = \omega_J + i\omega_K
\end{equation}
is a holomorphic symplectic form in complex structure $I$.  Thus we
can regard $X$ as a complex symplectic manifold with complex
symplectic structure $(I, \Omega_I)$, and construct Rozansky-Witten
theory, choosing $\nabla$ to be the Levi-Civita connection of $g$.

Of course, which complex structure to call $I$ is just a matter of
convention, and any other complex structure in the $\P^1$ gives an
equally good target space.  In other words, there is a family of
target spaces parametrized by the $\P^1$.  We can continuously change
the theory by moving within this family.  Equivalently, we may fix the
target space and vary the BRST operator.  In the hyperk\"ahler case
the theory has a second supercharge $\Qb$ which generates the
transformations
\begin{equation}
  \label{Qb-RW}
  \begin{alignedat}{2}
    \deltab\phi^i &= \Omega^{ij} g_{j\kb} \eta^\kb, &\qquad
    \deltab\phib^\ib &= 0,
    \\
    \deltab\chi^i
    &= -\Omega^{ij} g_{j\kb} \rmd\phib^\kb
       - \Gamma^i_{kj} \Omega^{kl} g_{l\mb} \eta^\mb \chi^j, &
    \deltab\eta^\ib &= 0,
  \end{alignedat}
\end{equation}
where $\Omega^{ij} \Omega_{jk} = \delta^i_k$.  As $\Qb$ squares to
zero and commutes with $Q$, any linear combination $Q_\zeta \propto Q
+ \zeta\Qb$ with $\zeta \in \P^1$ serves as a BRST operator.  (The
$Q$-exact part $S_1$ of the action is also $Q_\zeta$-exact since
$g_{i\jb} \chi^i \wedge \star\rmd\phib^\jb = \deltab(\Omega_{ij}
\chi^i \wedge \star\chi^j/2)$ and $\delta\deltab = (\delta +
\zeta\deltab) \deltab$.  Thus, the topological invariance of the
theory remains to hold.)  We see that for $\zeta \neq 0$, $Q_\zeta$
annihilates holomorphic functions on $X$ in a complex structure
different from $I$, so varying the BRST operator indeed amounts to
changing the complex structure of the target space.

\subsection[\texorpdfstring{$\Omega$}{Omega}-deformation of
  Rozansky-Witten theory]{\texorpdfstring{$\boldsymbol \Omega$}{Omega}-deformation of
  Rozansky-Witten theory}
\label{ORW}

Now let us formulate the $\Omega$-deformation of Rozansky-Witten
theory.  Our goal is the following.  Consider Rozansky-Witten theory
on $M = \R \times \Sigma$, equipped with a product metric $h = h_\R
\oplus h_\Sigma$.  Assume that the target space $X$ is a hyperk\"ahler
manifold with hyperk\"ahler metric $g$.  Given a vector field $V$
generating an isometry of $\Sigma$, we wish to construct a deformation
of this theory such that it has a supercharge $Q$ obeying the deformed
relation~\eqref{Q2LV}.

There are a couple of indications that such a deformation does exist.
One is that, as we will explain in section~\ref{RW4d},
Rozansky-Witten theory with hyperk\"ahler target space arises
naturally from $\CN = 2$ supersymmetric gauge theories in four
dimensions by compactification on $S^1$.  If we turn on an
$\Omega$-deformation in four dimensions, some deformation should be
induced in three dimensions as well.  Another indication is that
reduction of Rozansky-Witten theory on $S^1$ gives the B-model with
the same target space \cite{Thompson:1998vx}.  (More generally, if
the target space is not hyperk\"ahler, the dimensional reduction
yields a generalization of the B-model.)  As we could construct the
$\Omega$-deformation in two dimensions, it is natural to expect that
there is a corresponding deformation in three dimensions.  We take the
second observation as a starting point of our construction.

Our strategy is to describe the $\Omega$-deformed Rozansky-Witten
theory on $\R \times \Sigma$ with target space $X$ as an
$\Omega$-deformed B-twisted Landau-Ginzburg model on $\Sigma$ with
target space $Y = \Map(\R, X)$, the space of maps from $\R$ to $X$.
In order to specify the latter theory, we need to pick a complex
structure on $Y$.  Such a complex structure is naturally induced from
a complex structure chosen on $X$.  This construction therefore
singles out a distinguished element in the $\P^1$ of complex
structures on $X$.  We call it $I$.

Roughly speaking, having $\Map(\R, X)$ as the target space means that
we should regard a coordinate $t$ of the $\R$ as a continuous index,
putting it on the same footing as coordinate indices of $X$.  Hence,
the formula for the action \eqref{S0} now contains an integration over
$t$ in addition to summation over the other indices:
\begin{equation}
  \label{S0-RW}
  S_0
  = 
  \delta \int_{\R \times \Sigma} \sqrt{h_\R} \rmd t \wedge
  \Bigl(g_{i\jb}\rho^i
  \wedge \star_\Sigma\bigl(\rmd_\Sigma\phib^\jb + \iota_V\star_\Sigma\Fb^\jb\bigr)
  + g_{i\jb} F^i \wedge \star_\Sigma\mu^\jb\Bigr).
\end{equation}
Here $\star_\Sigma$ and $\rmd_\Sigma$ denote the Hodge star operator
and the exterior derivative (coupled to the Levi-Civita connection) on
$\Sigma$.  The supersymmetry transformation laws take the same
form~\eqref{SUSY} as before, the only difference being that the fields
have dependence on~$t$.

The above action lacks terms involving $t$-derivatives, which are
crucial for fully three-dimensional dynamics.  These missing terms are
to be provided by a superpotential $W$.  In the present context, $W$
is a holomorphic function on $\Map(\R, X)$ that respects locality,
namely a holomorphic functional of a map from $\R$ to $X$.

To construct a suitable superpotential, we complete the complex
structure $I$ into a triple $(I, J, K)$ compatible with the
hyperk\"ahler structure, and set $\Omega = \Omega_I$.  Since $\Omega$
is a closed holomorphic two-form, we can locally write $\Omega =
\rmd\Lambda$ with some holomorphic one-form $\Lambda$.  Using this
form we define $W$ by
\begin{equation}
  \label{W-RW}
  W(\Phi) = \frac12 \int_\R \Phi^*\Lambda.
\end{equation}
In this formula we have abused the notation and let $\Phi$ denote a
point on the target space $\Map(\R, X)$, as is customary in the
finite-dimensional setting.  With this choice of $W$, the
superpotential terms \eqref{SW} are given by
\begin{equation}
  \label{SW-RW}
  S_W
  = \frac{i}{2} \int_{\R \times \Sigma}
    \Bigl(\Omega_{ij} F^i \rmd_\R\phi^j
    - \frac{1}{2} \Omega_{ij} \rho^i \wedge \rmd_\R\rho^j
    + \Omegab_{\ib\jb} \Fb^\ib \rmd_\R\phib^\jb
    + \Omegab_{\ib\jb} \eta^\ib \rmd_\R\mu^\jb\Bigr).
\end{equation}

When $V = 0$, the theory described by the action $S_0 + S_W$ reduces
on-shell to Rozansky-Witten theory.  Integrating out the auxiliary
fields sets
\begin{equation}
  \sqrt{h_\R} \star_\Sigma F^i
  = -\frac{i}{2} g^{i\jb} \Omegab_{\jb\kb} \del_t\phib^\kb,
  \quad
  \sqrt{h_\R} \star_\Sigma\Fb^\ib
  = -\frac{i}{2} g^{\ib j} \Omega_{jk} \del_t\phi^k. 
\end{equation}
From the expression of $\Fb^\ib$, one readily sees that for $V = 0$,
the supersymmetry transformation laws \eqref{SUSY} coincide with the
formula \eqref{Q-RW}  under the identification
\begin{equation}
  \rho^i_\mu = \chi^i_\mu, \quad
  \sqrt{h_\R} \star_\Sigma\mu^\ib
  = -\frac{i}{2} g^{\ib j} \Omega_{jk} \chi^k_t .
\end{equation}
One can also check that the action coincides with the Rozansky-Witten
action under this identification.%
\footnote{In checking this, one uses the identity $g^{i\jb}
  \Omega_{ik} \Omegab_{\jb\lb} = 4g_{k\lb}$, which follows from the
  equation $\Omega = \omega_J + i\omega_K = -g(J + iK)$, and the fact
  that $\Omega$ is covariantly constant, which in particular implies
  $\Omega_{ik} R^k{}_{jl\mb} + \Omega_{kj} R^k{}_{il\mb} = 0$.}
For example, the F-term potential is
\begin{equation}
  \|\delta W\|^2
  = \frac14 \int_\R \rmd t \sqrt{h_\R} h^{tt}
    g^{i\jb} \Omega_{ik} \del_t\phi^k
    \Omegab_{\jb\lb} \del_t\phib^\lb
  = \int_\R \rmd t \sqrt{h_\R}
    g_{i\jb} h^{tt} \del_t\phi^i \del_t\phib^\jb,
\end{equation}
and this is precisely the kinetic term for the bosonic field along the
$t$-direction.  (Matching of the fermionic terms is straightforward.)
Thus, the theory constructed above provides a good definition for the
$\Omega$-deformation of Rozansky-Witten theory on $\R \times \Sigma$
with target space $X$ and complex symplectic structure $(I,
\Omega_I)$.

There is a slight generalization of this construction.  It is possible
to modify the definition of $W$ by terms that vanish for $V = 0$.
Locality requires that this is done through a deformation of $\Lambda$
by a locally-defined holomorphic one-form on $X$, which in turn gives
rise to a deformation of $\Omega$ by the equation $\Omega =
\rmd\Lambda$.  The latter is what really matters as far as the effect
on the theory is concerned.  After a deformation of this type is
included, generically $\Omega$ would remain nondegenerate and define a
deformed complex symplectic structure.  However, it may no longer be a
holomorphic symplectic form associated to some complex structure in a
hyperk\"ahler structure.  This generalization will not be considered
in what follows, except that we will briefly discuss its relevance in
applications to $\CN = 2$ supersymmetric gauge theories in four
dimensions.

\subsection{Reduction to quantum mechanics}

Finally we are ready to present the main result of this paper.
Suppose that $\Sigma$ is a disk $D$ and $V$ generates its rotations.
In this situation we can apply the localization formula for
correlation functions obtained in the previous section.  Using the
formula, we show that this system is equivalent to a quantum
mechanical system on a real symplectic submanifold of $X$.

First of all, we have to specify the support $\gamma$ of the brane
placed on the boundary of $D$.  We recall that $\gamma$ is a
Lagrangian submanifold of $Y = \Map(\R, X)$, and $\Im W$ must be
locally constant on $\gamma$.  The first condition suggests that we
should take $\gamma = \Map(\R, L)$, with $L$ being a Lagrangian
submanifold of $X$ with respect to the K\"ahler form $\omega_I$.  The
second condition says that we must have
\begin{equation}
  \delta\Im W
  = \frac12 \int_\R \Im\Omega_{ij} \delta\phi^i \rmd\phi^j
  = 0
\end{equation}
on the boundary.  This is satisfied if $L$ is Lagrangian with respect
to $\Im\Omega = \omega_K$.  Then, both $I$ and $K$ give an isomorphism
between $T_\R L$ and $N_\R L$, and $J = KI$ is an endomorphism of
$T_\R L$.  Hence, $L$ is a Lagrangian submanifold with respect to
$\omega_I$ and $\omega_K$, while a complex submanifold in $J$.  In
particular, it is a symplectic manifold with symplectic form
$\Re\Omega = \omega_J$.

We also need to specify the locally constant function $W_0$ on
$\gamma$, which is part of the boundary term.  For this one, we
introduce a $\U(1)$ connection $A$ on a line bundle over $L$ and set
\begin{equation}
  \label{W0-RW}
  W_0 = \frac12 \int_\R \Phi^*A.
\end{equation}
For $W_0$ to be locally constant, $A$ must be flat.

For simplicity, let us assume for a moment that $\Omega$ is an exact
form so that a holomorphic one-form $\Lambda$ satisfying $\Omega =
\rmd\Lambda$ exists globally on $X$.  Then $W$ is given by the
integral \eqref{W-RW}, and the localization formula \eqref{LF} reads
\begin{equation}
  \vev{\CO}
  = \int_{\Map(\R, L)} \cD\Phi_0
    \exp\Bigl(\frac{i}{\hbar} S(\Phi_0)\Bigr) \CO(\Phi_0),
\end{equation}
where the action $S$ and the Planck constant $\hbar$ are given by
\begin{equation}
  \label{SQM}
  S(\Phi_0) = \int_\R \Phi_0^*(\Re\Lambda + A), \quad
  \hbar = \frac{\veps}{\pi}.
\end{equation}
The right-hand side of the formula is the path integral for a quantum
mechanical system with phase space $(L, \Re\Omega)$; in local Darboux
coordinates $(p_a, q^a)$, $a = 1$, $\dotsb$, $\frac12\dim L$ such that
$\Re\Omega|_L = \rmd p_a \wedge \rmd q^a$, we have $\Re\Lambda|_L + A
= p_a \rmd q^a$ up to an exact form, so the Lagrangian is $p_a \dot
q^a$ up to a total derivative.  Therefore, this system quantizes the
symplectic manifold $(L, \Re\Omega)$.  Notice that the Hamiltonian of
the system is zero, as is consistent with the fact that we started
with a (quasi-)topological field theory.

Observables of the $\Omega$-deformed Rozansky-Witten theory include
local operators that may be regarded as elements of the
$\delb$-cohomology $H^{0,\bullet}(X; \C)$ in complex structure $I$,
inserted at the origin of $D$ and arbitrary points on the $\R$.  Among
these observables, those that are nonvanishing after the localization
are holomorphic functions on $X$.  The path integral turns these
observables into operators in the quantum mechanical system which form
a noncommutative algebra.  They act on the Hilbert space whose
elements are, say, functions of $q^a$ locally.  What we have obtained
is thus a noncommutative deformation of the algebra of functions on
$L$ that are restrictions of holomorphic functions on $X$, acting on
the space of sections of a hermitian line bundle over $L$.

It is worth noting that the localization formula derived here is very
similar to one for an $\CN = 4$ supersymmetric sigma model on $\R
\times S^2$, constructed from twisted chiral multiplets of $\CN =
(2,2)$ supersymmetry on $S^2$ \cite{Luo:2013nxa}.  In that case, the
action contains the term
\begin{equation}
  \frac{2i}{r} \int_{S^2} \Re W,
\end{equation}
where $r$ is the radius of the $S^2$ \cite{Gomis:2012wy}.  This term
corresponds to our boundary term \eqref{BT-2}, and eventually becomes
the action of a quantum mechanical system.  The localization again
requires the bosonic field to be constant, so the above term is
multiplied by the area of the $S^2$ and the Planck constant is
proportional to $1/r$.

Let us discuss what changes have to be made if we remove the
assumption that $\Omega$ is exact.  In this case $\Lambda$ can exist
only locally, so the formula for $W$ is not well-defined.  This is not
a problem when we work with a spacetime with no boundary, since what
enters the action then is not $W$ itself, but the derivative of $W$,
which can be expressed in terms of $\Omega$.  However, it does cause a
problem in the present setup where $W$ enters the boundary term in the
action.

A better definition for $W$ is the following.  First, in each homotopy
class $\CP$ of $\Map(\R,X)$ we fix a reference map $\Phi_\CP$, so that
any map $\Phi \in \CP$ can be deformed to $\Phi_\CP$ without \linebreak altering
the behavior at infinity.  Next, given a map $\Phi \in \CP$, we pick a
homotopy $\Phih\colon [0,1] \to$ $\Map(\R,X)$ from $\Phi_\CP$ to
$\Phi$; thus $\Phih(0) = \Phi_\CP$ and $\Phih(1) = \Phi$.  Finally, we
define
\begin{equation}
  W(\Phi) = \frac12 \int_{[0,1] \times \R} \Phih^*\Omega,
\end{equation}
viewing $\Phih$ as a map from $[0,1] \times \R$ to $X$.  Changing the
reference maps $\Phi_\CP$ shifts $W$ by a locally constant function,
but such a shift can be absorbed in the definition of $W_0$.  If
$\Omega$ is exact, $W$ is given as before by the integral of a
holomorphic one-form $\Lambda$ such that $\Omega = \rmd\Lambda$.
Since the functional derivative of $W$ can be computed locally, and
locally we can always write $\Omega$ as $\Omega = \rmd\Lambda$, this
definition of $W$ leads to the same superpotential terms
\eqref{SW-RW}.

The function $W$ so defined is actually not single-valued on $\Map(\R,
X)$.  If we choose a different homotopy $\Phih'$, then the two
homotopies combine into a map $\Delta\Phih\colon S^1 \times \R \to X$,
and $W$ changes by
\begin{equation}
  \Delta W = \frac12 \int_{S^1 \times \R} \Delta\Phih^*\Omega.
\end{equation}
By assumption the homotopies leave the behavior at infinity intact, so
$\Delta\Phih$ maps each end of the cylinder $S^1 \times \R$ to a
point.  As such, $\Delta\Phih$ may be thought of as really a map from
a two-sphere to $X$.  For the path integral with the boundary term
\eqref{BT-2} to be well-defined, the change in the boundary term must be
always an integer multiple of $2\pi i$.  This requirement places the
constraint
\begin{equation}
  \frac{1}{2\pi\hbar} [\Re\Omega] \in H^2(L; \Z).
\end{equation}
This is nothing but the quantization condition for $(L,\Re\Omega)$.

Now we summarize what we have found.  Let $(X, g, I, J, K)$ be a
hyperk\"ahler manifold.  Pick a submanifold $L$ of $X$ that is
Lagrangian with respect to $\omega_I$ and $\omega_K$, and holomorphic
in $J$.  Then, the $\Omega$-deformed Rozansky-Witten theory with
target space $X$ in complex symplectic structure $(I, \Omega_I)$,
formulated on $\R \times D$ with boundary condition specified by a
brane supported on $L$, is equivalent to a quantum mechanical system
whose phase space is the symplectic manifold $(L, \omega_J)$.  The
Planck constant is proportional to the $\Omega$-deformation parameter
$\veps$.

\subsection{Comparison with the A-model approach}
\label{AA}

In \cite{Gukov:2008ve}, Gukov and Witten developed a framework for
quantization of symplectic manifolds using branes in the A-model.  It
is illuminating to compare their approach with ours.

In the A-model approach, one first embeds the symplectic manifold
$(L,\omega)$ that one wants to quantize into a complex symplectic
manifold $(X, \Omega)$ of twice the dimension such that $\Re\Omega|_L
= \omega$ and $\Im\Omega|_L = 0$.  (We are using the same symbols as
in our approach to emphasize parallels.)  One then considers the
A-model whose target space is the symplectic manifold $(X,
\Im\Omega)$.  Taking the worldsheet to be a strip, one places two
types of A-branes on its sides.  On one side is a Lagrangian A-brane
supported on $L$ and endowed with a complex line bundle with a flat
$\U(1)$ connection $A$.  On the other is a canonical coisotropic
A-brane \cite{Kapustin:2001ij} whose support is the whole $X$, endowed
with a line bundle with a connection of curvature $\Re\Omega$.  It
turns out that this system quantizes $(L, \omega)$ just like our
system does: the open strings with both ends attached on the canonical
coisotropic brane form a noncommutative deformation of the algebra of
holomorphic functions on $X$, whereas the strings stretched between
the two branes span a Hilbert space on which the deformed algebra
acts.

A particularly nice situation is when $X$ admits a hyperk\"ahler
structure such that $\Omega = \Omega_I$ and $L$ is a complex
submanifold in complex structure $J$, since in this case one can study
the B-model of complex structure $J$ and describe the Hilbert space
explicitly.  Since $\omega_K = \Im\Omega$ vanishes on $L$, so does
$\omega_I = -\omega_K J$ then.  Thus $L$ is a Lagrangian submanifold
with respect to $\omega_I$ and $\omega_K$, and a complex submanifold
in complex structure $J$; the branes are of type $(A,B,A)$.
Interestingly, this is precisely the property required of the support
of a brane used in our approach.  In our case we have a single brane
on the boundary of $D$, and it may be thought of as playing the role
of a combination of the two branes in the A-model approach.  For
example, $\Re W$ and $W_0$ correspond to the gauge fields on the
space-filling and middle-dimensional branes, respectively, as can be
seen from their expressions~\eqref{W-RW} and~\eqref{W0-RW}.

In view of these similarities, it may be reasonable to expect that if
we equip $D$ with a cigar metric and regard $D$ as an $S^1$-fibration
over an interval, our system reduces to the A-brane system at low
energies.  Nekrasov and Witten \cite{Nekrasov:2010ka} showed that this
is indeed the case under certain circumstances.  We will discuss this
point briefly in section \ref{QO4d}

\section{Applications to four-dimensional gauge theory}
\label{4d}

In the final section we discuss applications of our framework to $\CN
= 2$ supersymmetric gauge theories in four dimensions.  For each of
these theories, there is a class of physical quantities captured by
Rozansky-Witten theory with hyperk\"ahler target space.  Using this
fact and the results obtained in the previous section, we establish
connections between the gauge theory and quantization of objects
associated with the target space.

\subsection{Rozansky-Witten theory from four dimensions}
\label{RW4d}

\enlargethispage{9.7pt}
 
First we clarify the relation between $\CN = 2$ supersymmetric gauge
theories and Rozansky-Witten theory, and explain some important
features of the geometry of the emergent target spaces.  For more
details, see \cite{Seiberg:1996nz, Gaiotto:2008cd}.

Consider an $\CN = 2$ supersymmetric gauge theory, compactified on
$S^1$.  On the Coulomb branch, the theory is described at low energies
by a three-dimensional abelian gauge theory with $\CN = 4$
supersymmetry.  Dualizing the gauge fields to periodic scalars, we get
a sigma model.  Its target space $\CM$ is required to be hyperk\"ahler
by the $\CN = 4$ supersymmetry, and has dimension $4r$, where $r$ is
the rank of the gauge group.  For a large class of theories obtained
by compactification of M5-branes on punctured Riemann surfaces, $\CM$
is the Hitchin moduli space of the relevant surface
\cite{Gaiotto:2009we, Gaiotto:2008cd}.

If we instead start from the topologically twisted version of the same
theory, then the resulting sigma model is twisted as well. Placing the
ultraviolet theory on $M \times S^1$, we get Rozansky-Witten theory
on $M$ with target space $\CM$.  Recall that when the target space is
hyperk\"ahler, Rozansky-Witten theory has two supercharges $Q$ and
$\Qb$.  The first one exists in the more general case of complex
symplectic target spaces, but the second does not.  From the
four-dimensional viewpoint, $Q$ is the scalar supercharge of the
twisted theory, while $\Qb$ is the component of the one-form
supercharge along the $S^1$.  Any linear combination $Q_\zeta \propto
Q + \zeta\Qb$ with $\zeta \in \P^1$ may be used as a BRST operator.
This corresponds to the fact that $\CM$ has a $\P^1$-worth of complex
structures $J_\zeta$ in which Rozansky-Witten theory can be
formulated.  We write $\Omega_\zeta$ for the holomorphic symplectic
form associated to $J_\zeta$.

The above effective description is valid at length scales that are
much larger than the radius of the $S^1$ so that the theory looks
effectively three-dimensional, but much smaller compared to the size
of $M$ so that the effects of the curvature of $M$ are negligible on
the massive modes that are integrated out.  This requirement can
always be met by rescaling of the spacetime metric which leaves
physical quantities unaffected.  This is clearly true for $\zeta = 0$,
that is when the BRST operator is $Q$, in which case the twisted
theory is well-known to be a topological field theory
\cite{Witten:1988ze}.  It is also true for $\zeta \neq 0$.  The reason
is that correlation functions of $Q_\zeta$-invariant operators on $M
\times S^1$ are supersymmetric indices and protected under
deformations of the parameters of the theory.%
\footnote{We assume that the $Q_\zeta$-invariant states form a
  discrete spectrum, which should be the case if $M$ is compact.
  Although the choice $M = \R \times D$ that we will consider is not
  compact, we can replace the $\R$ by a finite interval without
  altering the conclusions.}

The geometry of $\CM$ is very interesting, in that there is a
distinguished complex structure in which $\CM$ is the phase space of a
complex integrable system \cite{Donagi:1995cf}.  This is the complex
structure $J_0$, and usually called $I$.  In this complex structure,
$\CM$ is a torus fibration over a complex manifold $\CB$ whose fibers
are complex Lagrangian submanifolds with respect to the holomorphic
symplectic form $\Omega_I$.  The base $\CB$ is the Coulomb moduli
space of the ultraviolet theory placed on $\R^4$; it is topologically
an affine space $\C^r$, parametrized by the vacuum expectation values
of the gauge-invariant polynomials in the vector multiplet scalar.
The torus fibers are parametrized by the holonomies of the infrared
abelian gauge fields and their magnetic duals around the $S^1$.

There are particularly nice coordinates on $\CM$ in this context.
$\CN = 2$ supersymmetry requires that $\CB$ admits local holomorphic
coordinates $a^i$, $i = 1$, $\dotsb$, $r$, and their duals $a_{D,i}$
related through a holomorphic function $\CF$ as $a_{D,i} =
\del\CF/\del a^i$.  The second derivatives $\tau_{ij} = \del^2\CF/\del
a^i \del a^j = \del a_{D,i}/\del a^j$ encode the complexified gauge
couplings of the effective abelian gauge theory.  If we write the
electric and magnetic holonomies as $\exp(i\theta_e^i)$ and
$\exp(i\theta_{m,i})$, then $z_i = \theta_{m,i} - \tau_{ij}
\theta_e^j$ are holomorphic coordinates on the torus fiber of $\CM$.
Moreover, $(a^i, z_i)$ are complex Darboux coordinates:%
\footnote{Holomorphic objects in complex structure $I$, especially the
  structure of complex integrable system, do not receive instanton
  corrections coming from BPS particles circling around the $S^1$.  This is
  because the action for such particles is not holomorphic, but rather
  the absolute value of a holomorphic function \cite{Seiberg:1996nz}.}
\begin{equation}
  \Omega_I = \rmd a^i \wedge \rmd z_i.
\end{equation}
The $a^i$ are conserved charges generating translations in the fiber
directions, and commute with one another with respect to the Poisson
bracket derived from $\Omega_I$.  There are $r$ such charges in the
phase space $\CM$ of complex dimension $2r$, reflecting the fact that
the system is completely integrable in the complex sense.

\subsection{Quantization by twisting of the spacetime}

Now we wish to modify the ultraviolet theory in such a way that the
effective theory undergoes an $\Omega$-deformation.  For $\zeta \neq
0$, $\infty$, we can achieve this by replacing the spacetime $M \times
S^1$ with a twisted product between $M$ and $S^1$, which is a
nontrivial $M$-fibration over $S^1$.  More specifically, we take the
trivial fibration $M \times [0,1]$, and identify the fibers at the two
ends of the interval $[0,1]$ with the action of an isometry of $M$.
Writing this isometry as $\exp(V)$ with $V$ a Killing vector field, we
denote the resulting fibration by $M \times_V S^1$.

Since $Q_\zeta$ are scalars on $M$, it commutes with isometries on $M$
and correlation functions of $Q_\zeta$-invariant operators on $M
\times_V S^1$ are still protected indices.  As such, they may be
computed by the effective sigma model.  We have $\{Q, \Qb\} \propto
P_4$ and hence $Q_\zeta^2 \propto P_4$ for $\zeta \neq 0$, $\infty$
(up to a central charge), where $P_4$ acts on fields by $\del_4$.  Due
to the isometry twist, at low energies $P_4$ is replaced by $L_V$,%
\footnote{To see this, one can use coordinates $(y^\mu, y^4) =
  (\exp(x^4 V) x^\mu, x^4)$, where $x^\mu$ and $x^4$ are coordinates
  on $M$ and $S^1$, respectively.  In these coordinates the fibration
  is ``untwisted,'' $(y^\mu, 0) \sim (y^\mu, 1)$, and at low energies
  we simply have $\del/\del y^4 = \del/\del x^4 - V = 0$ on
  functions.}
leading to the deformed relation $Q_\zeta^2 \propto L_V$ in three
dimensions.  We thus identify the effective theory for $\zeta \neq 0$,
$\infty$ with Rozansky-Witten theory subject to the
$\Omega$-deformation by a (complex) Killing vector field proportional
to $V$.

Taking $M = \R \times D$ and $V$ to generate rotations of the disk
$D$, we conclude that a twisted $\CN = 2$ supersymmetric gauge theory
on the corresponding fibration quantizes a symplectic submanifold $(L,
\Re\Omega_\zeta)$ of $\CM$, where $L$ is the support of the brane on
the boundary of $D$.

It is interesting to consider $Q_\zeta$-invariant operators.  In three
dimensions, relevant operators are holomorphic functions on $\CM$ in
complex structure $J_\zeta$, inserted at the center of $D$ and points
on the $\R$.  They form a noncommutative deformation of the algebra of
holomorphic functions on $\CM$.  In four dimensions, these local
observables may be represented by line operators wrapped on the $S^1$.
This explains the observation made in \cite{Gaiotto:2010be,
  Ito:2011ea} that supersymmetric loop operators realize a deformation
quantization of the algebra of holomorphic functions on $\CM$.

We remark that due to the twisting of the spacetime, there may be
corrections to the holomorphic symplectic form $\Omega_\zeta$ when it
appears in the $\Omega$-deformed Rozansky-Witten theory and hence in
the quantum mechanical system; see the comment at the end of
section~\ref{ORW} for this point.

\subsection[Quantization by the \texorpdfstring{$\Omega$}{Omega}-deformation in four dimensions]{Quantization by the \texorpdfstring{$\boldsymbol \Omega$}{Omega}-deformation in four dimensions}
\label{QO4d}

The above argument does not apply when $\zeta = 0$ or $\infty$.  For
$\zeta = 0$, there is a more obvious way to induce an
$\Omega$-deformation in the effective Rozansky-Witten theory.  That
is to turn on an $\Omega$-deformation in the ultraviolet.

What we find in this case is that a twisted $\CN = 2$ supersymmetric
gauge theory on $\R \times D \times S^1$, subject to the
$\Omega$-deformation by a rotation generator of $D$, quantizes a real
symplectic submanifold $(L, \Re\Omega_I)$ of the complex integrable
system $(\CM, \Omega_I)$. The commuting Hamiltonians of the quantum
integrable system are the operators corresponding to the special
coordinates $a^i$.  In the ultraviolet theory, they are realized
by the gauge-invariant polynomials in the vector multiplet scalar
inserted at the origin of $D$.

Since the twisted theory is topological and its Hamiltonian is
identically zero, the states of the quantum mechanical system
correspond to the vacua of the gauge theory.  Let us find where the
vacua are located.  To be concrete, we take $L$ to be the locus given
locally by the equations
\begin{equation}
  \Im a_{D,i} = \theta_{m,i} = 0.
\end{equation}
This is a good choice; we may choose a hyperk\"ahler metric such that
$\omega_I|_L = 0$ (for example, the semiflat metric obtained by
dimensional reduction of the effective abelian gauge theory), while
$\Omega = \rmd a^i \wedge \rmd\theta_{m,i} - \rmd a_{D,i} \wedge
\rmd\theta_e^i$ and hence $\Im\Omega_K|_L = 0$, showing that $L$ is a
Lagrangian submanifold with respect to $\omega_I$ and $\omega_K =
\Im\Omega$ as required.  $L$ is a real integrable system over the real
submanifold of $\CB$ parametrized by $\Re a_{D,i}$, with the torus
fiber parametrized by $\theta_e^i$.  The Lagrangian of the quantum
mechanical system is $-\Re a_{D,i} \rmd\theta_e^i$.  Integrating over
the periodic scalars $\theta_e^i$ imposes the constraints $\Re
a_{D,i}/\hbar \in \Z$.  Combining these constraints with the equations
$\Im a_{D,i} = 0$, we obtain
\begin{equation}
  \exp\Bigl(\frac{2\pi i}{\hbar} a_{D,i}\Bigr) = 1.
\end{equation}
These are the equations that determine the locations of the vacua.

The above equations are to be identified with the Bethe equations in
the integrable system \cite{Nekrasov:2009rc}.  Let $\CF$ be the
($\veps$-corrected) prepotential of the gauge theory, and $\Wt = 2\pi
i\CF/\hbar$.  Then we can rewrite the equations as
\begin{equation}
  \exp\Bigl(\frac{\del\Wt}{\del a^i}\Bigr) = 1.
\end{equation}
If the radius of $S^1$ is much larger than $1/\veps$, then there is a
low-energy regime in which the $\Omega$-deformed theory is effectively
described by a two-dimensional theory with $\CN = (2,2)$
supersymmetry.  The function $\Wt$ may be interpreted as the twisted
superpotential for this effective theory on $\R \times S^1$.  In the
quantum integrable system, it is interpreted as the Yang-Yang
function.

In the work of Nekrasov and Witten \cite{Nekrasov:2010ka}, the
quantization of the integrable system was explained in the A-brane
framework disscussed in section \ref{AA}.  The starting point of their
approach is the same as ours, namely the twisted theory on $\R \times
D \times S^1$, subject to the $\Omega$-deformation on $D$.  One puts a
cigar metric on $D$ and thinks of it as an $S^1$-fibration over an
interval.  One then reduces the theory to a two-dimensional theory on
a strip.  It turns out that away from the tip of the cigar, the effect
of the $\Omega$-deformation can be canceled by a redefinition of
fields.  One makes use of this observation and deduces that the
two-dimensional theory is an $\CN = (4,4)$ supersymmetric sigma model
with target space $\CM$, with a space-filling $(A, B, A)$-brane placed
on one side of the strip and a middle-dimensional $(A,B,A)$-brane
placed on the other.  This configuration fits in the A-brane
framework, and one concludes that it quantizes a real integrable
system which is the support of the middle-dimensional brane.  Here we
have presented an alternative derivation based on the framework
developed in this paper.  The relation between the two approaches
should be clear from the discussion in section \ref{AA}.

\section*{Acknowledgments}

I would like to thank Yuan Luo for helpful discussions, and Petr Vasko
for comments on the manuscript.  This work is supported by INFN
Postdoctoral Fellowship and INFN Research Project ST\&FI.

\appendix

\section{\texorpdfstring{$\boldsymbol \Omega$}{Omega}-deformation of twisted \texorpdfstring{$\boldsymbol{\CN
  = 2}$}{N = 2} supersymmetric gauge theories}
\label{Omega}

In this appendix we review the $\Omega$-deformation of topologically
twisted $\CN = 2$ supersymmetric gauge theories in four dimensions
\cite{Nekrasov:2002qd, Nekrasov:2003rj}.  We only consider the case of
pure super Yang-Mills theory, constructed from a vector multiplet.
The bosonic fields of the twisted theory are a gauge field $A$, a
complex scalar $\phi$, and an auxiliary self-dual two-form $H$.  The
fermionic fields are a zero-form $\eta$, a one-form $\psi$, and a
self-dual two-form $\chi$.  The spacetime can be any four-manifold $M$
that admits an isometry.

To introduce an $\Omega$-deformation, one lifts the theory to a
six-dimensional gauge theory, formulated on a nontrivial $M$-fibration
over a two-torus $T^2$ such that the fiber is acted upon by isometries
as one goes around cycles in the base.  Killing vector fields
generating the isometries are assumed to commute with each other.  One
then dimensionally reduces the lifted theory down to four dimensions.
Formally this procedure has the effect of replacing $\phi$ and its
hermitian conjugate $\phib$ by differential operators as $\phi \to
\phi + V^\mu D_\mu$ and $\phib \to \phib + \Vb^\mu D_\mu$, where $V$
is a linear combination of the Killing vector fields and $\Vb$ is its
complex conjugate.

The $\Omega$-deformed supersymmetry transformation laws are
\begin{equation}
  \label {SUSY-4d}
  \begin{aligned}
    \delta A &= \psi,
    \\
    \delta\phi &= \iota_V\psi,
    \\
    \delta\phib &= \eta + \iota_{\Vb}\psi,
    \\
    \delta\eta
    &= i[\phi, \phib] - \iota_\Vb \rmd_A\phi + \iota_V \rmd_A\phib
         + \iota_V \iota_{\Vb} F_A,
    \\
    \delta\psi &= \rmd_A\phi + \iota_V F_A,
    \\
    \delta\chi &= iH,
    \\
    \delta H  &= [\phi, \chi] - i\CL_V\chi.
  \end{aligned}
\end{equation}
Here $\rmd_A = \rmd - iA$ is the exterior derivative coupled to $A$,
and $F_A$ is the curvature of $A$.  The above transformation preserves
the self-duality of $H$ since the (gauge-covariant) Lie derivative
$\CL_V = \rmd_A\iota_V + \iota_V\rmd_A$ commutes with the Hodge
duality if $V$ is a Killing vector field.  The supersymmetry algebra
closes provided that $V$ and $\Vb$ commute.  The generator $Q$ of the
supersymmetry satisfies $Q^2 = L_V$ modulo a gauge transformation,
where $L_V$ is the conserved charge that acts on fields as $\CL_V$.

The action of the $\Omega$-deformed theory is
\begin{equation}
  S = \frac{\Im\tau}{4\pi} \delta \int_M \Tr\Bigl(
        \frac12 \bigl(-\delta\chi + 4F_A^+\bigr) \wedge \star\chi
        + \overline{\delta\psi} \wedge \star\psi
        + \frac12 \overline{\delta\eta} \wedge \star\eta\Bigr)
        + \frac{i\tau}{4\pi} \int_M \Tr F_A \wedge F_A,
\end{equation}
where $\tau = \theta/2\pi + 4\pi i/e^2$ is the complexified gauge
coupling and $F_A^+$ is the self-dual part of~$F_A$.  Integrating out
the auxiliary field, we find that the bosonic part of the action is
given by
\begin{multline}
  \frac{\Im\tau}{4\pi} \int_M \Tr\Bigl(
  F_A \wedge \star F_A
  + \bigl(\rmd_A\phib + \iota_{\Vb} F_A\bigr)
    \wedge \star\bigl(\rmd_A\phi + \iota_V F_A\bigr) \\
  + \frac12 \bigl([\phi, \phib] + i\iota_\Vb\rmd_A \phi - i\iota_V\rmd_A\phib
  - i\iota_V \iota_{\Vb} F_A\bigr)^2\Bigr)
  + \frac{i\Re\tau}{4\pi} \int_M \Tr F_A \wedge F_A.
\end{multline}
When $V = 0$, this reduces to the bosonic part of the standard $\CN =
2$ super Yang-Mills action.

\bibliography{yagi}{}
\bibliographystyle{JHEP}

\end{document}